\pdfoutput=1
\documentclass{JFM-FLM_Au}

\usepackage{placeins}
\graphicspath{{figures/}}

\AtBeginDocument{\raggedbottom}

\lefttitle{J. Li and C. Luo}
\righttitle{Explicit wave-angle solutions for equilibrium oblique detonations}

\title{Exact explicit wave-angle solutions for equilibrium oblique detonations}

\author{Jing Li\aff{1,2} \and Changtong Luo\aff{1,2}}

\affiliation{%
\aff{1}State Key Laboratory of High Temperature Gas Dynamics,
Institute of Mechanics, Chinese Academy of Sciences,
Beijing 100190, China
\aff{2}School of Engineering Science,
University of Chinese Academy of Sciences,
Beijing 100049, China}

\corresau{Changtong Luo, \email{luo@imech.ac.cn}}

\begin{document}

\maketitle

\begin{abstract}
The equilibrium oblique-detonation polar is a fundamental gasdynamic relation.
This relation links the deflection angle, the wave angle, the flow state and the heat release.
However, the wave angle is usually determined by numerical iteration. 
There is no explicit formula to determine the wave angle from a prescribed deflection angle,
referred to as the inverse problem, 
even for the classical oblique-detonation model: a calorically perfect gas with a 
constant specific-heat ratio and a fixed heat release. 
This is inconvenient in practical applications.
In this work, we solve this inverse problem for the classical model 
and derive explicit formulae for the wave angle.
Inspired by the derivation of the oblique-shock inverse relation, 
the oblique-detonation inverse relation also reduces to a cubic equation. 
Heat release only changes the cubic coefficients, so the
equation can still be solved explicitly. We also derive the complete analytical branch structure.
We further find that the detachment
point and the downstream total-sonic point can be determined explicitly by
solving a cubic and a quadratic equation, respectively. 
We extend the explicit formulae to the two-$\gamma$ model
and apply them to oblique-detonation calculations
with equilibrium chemistry.
These explicit formulae for the wave angle and critical points
promote the theoretical understanding of oblique detonations
and support their practical calculation.
\end{abstract}

\section{Introduction}
\label{sec:introduction}

The equilibrium oblique-detonation polar is the basic gasdynamic relation for a
steady detonation attached to a wedge \citep{teng2020,jiang2023}. In the
classical model, the gas is calorically perfect with constant $\gamma$, and the
heat release $Q$ is prescribed. For fixed upstream conditions, prescribing the
wave angle $\beta$ determines the corresponding deflection angle $\theta$ and
downstream equilibrium state. Varying $\beta$ traces the polar; we refer to this
$\beta$-parameterised evaluation as the forward problem. Early detonation-polar analyses were given by
\citet{siestrunck1953}, \citet{gross1963} and \citet{townend1970}; the CJ and
branch structure was subsequently developed and reviewed
\citep{pratt1991,powers1994,powersstewart1992}.

In many practical applications, however, the deflection angle $\theta$ is prescribed by the wedge geometry,
while $M$, $\gamma$ and $Q$ follow from the upstream state and the thermochemical model, 
and we want to find the corresponding wave angle $\beta$.
The forward formulation does not allow us to calculate $\beta$ explicitly.
The wave angle is usually then read from plotted $\theta$-$\beta$-$M$-$Q$ curves,
or obtained by numerical iteration \citep{li2025criteria}.
This leads to the inverse problem studied here:
for given $(\theta,M,\gamma,Q)$, determine all algebraic wave-angle candidates
and identify the equilibrium branch and admissibility of each. 
The present study first considers the classical model and asks whether
the polar admits an exact, explicit algebraic inversion.

A useful reference is the inert oblique shock. For this problem,
the inverse $\theta$--$\beta$--$M$ relation reduces to a cubic equation,
written either in $\sin^{2}\beta$ or in $\tan\beta$ \citep{thompson1950,wellmann1972}.
The admissible wave angles can therefore be obtained in closed form
\citep{mascitti1969,wellmann1972,hartley1991,emanuel2001}.
The detonation polar has the same wedge geometry, but heat release changes its root structure.
It introduces the CJ bound and an additional low-compression equilibrium root. 
The weak and strong root classification used for inert shocks is therefore not sufficient. 
\citet{townend1970} derived explicit relations, including a detachment condition,
for families in which a wave parameter $F$ is held constant. However, $F$ varies along a 
fixed-heat-release polar. A constant-$F$ relation is therefore different from the fixed-$Q$ 
relation considered here.

In this work, we show that the fixed-$Q$ oblique-detonation polar can be
represented by a small set of low-order algebraic equations. For prescribed
$(\theta,M,\gamma,Q)$, the wave angles satisfy a cubic equation in
$t=\tan\beta$. All candidate roots can be obtained in closed form using the
same algebraic methods as for the classical oblique-shock cubic. An algebraic
sign criterion assigns each retained real root to the high- or low-compression
branch, making the inversion branch-complete.
In the limit $Q\to0$, the cubic reduces term by term to the classical oblique-shock equation.
The merging of the weak and strong roots gives a second cubic for the fixed-$Q$ detachment point.

We then determine the downstream state of each solution.
The downstream total-sonic condition reduces exactly to a quadratic equation in
\(u=M^2\sin^2\beta\).
For every attached high-compression polar, this quadratic equation has exactly one physical root.
The root lies between the CJ endpoint and the detachment point.
It divides the weak branch into a downstream-supersonic and a downstream-subsonic part.
The strong branch is subsonic downstream throughout.
The closed-form results are further extended to the two-$\gamma$
model through an exact parameter mapping.
The closed-form formulae can also be applied to chemical-equilibrium
calculations. We incorporate them into a two-step iterative framework
to replace the inner wave-angle iteration and reduce the computational
cost.
An open-source Python implementation of the proposed methods is provided.

\section{Exact inversion and admissible wave-angle solutions}
\label{sec:inversion}

\subsection{Model and classical detonation polar}
\label{sec:model}

The equilibrium model is as follows.
A uniform supersonic stream of a combustible gas mixture passes over a straight
wedge and supports a steady, planar, attached oblique detonation.
The flow is two-dimensional, and outside the front it is inviscid and adiabatic.
The gas is calorically perfect, with the same gas constant $R$ and
specific-heat ratio $\gamma>1$ on both sides.
Section~\ref{sec:two-gamma} relaxes this assumption and allows the
specific-heat ratio to differ on the two sides of the front.
Heat release is taken to be instantaneous and complete, so the reaction zone is
absorbed into the front and the state behind it is the fully reacted
equilibrium state.
The front is therefore a discontinuity of zero thickness, and only the local
Rankine--Hugoniot end states enter the analysis.

\begin{figure}[!ht]
 \centering
 \includegraphics[width=0.9\linewidth]{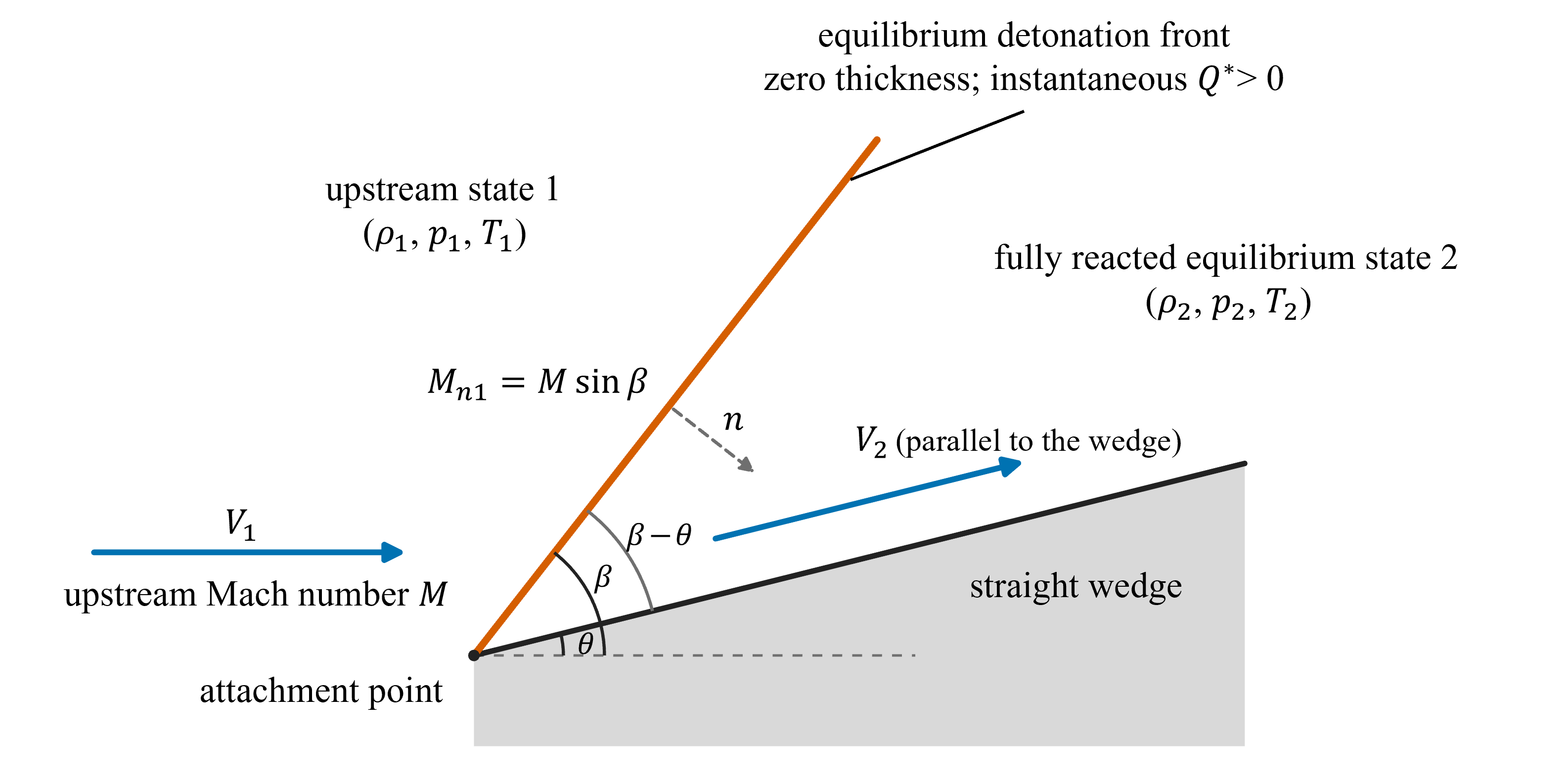}
 \caption{Geometry and notation for a planar equilibrium oblique detonation
 attached to a straight wedge. The front makes an angle $\beta$ with the
 upstream velocity, the flow is deflected through $\theta$, and
 $M_{n1}=M\sin\beta$ is the upstream normal Mach number.}
 \label{fig:oblique-detonation-geometry}
\end{figure}

Subscripts 1 and 2 denote the upstream state and the fully reacted downstream
equilibrium state, respectively. 
Let $M$ and $T_1$ denote the upstream Mach number and static temperature,
and let $\beta$ and $\theta$ denote the wave and flow-deflection angles,
respectively. For a chemical energy release $Q^*>0$ per unit mass, the
dimensionless heat release used here is $Q=Q^*/(RT_1)$. The geometric
configuration and angle definitions are summarised in
figure~\ref{fig:oblique-detonation-geometry}.

A classical representation of the equilibrium oblique-detonation polar
\citep{gross1963,pratt1991} is
\begin{equation}
 \frac{\tan\beta}{\tan(\beta-\theta)}
 =\frac{(\gamma+1)M^2\sin^2\beta}
 {1+\gamma M^2\sin^2\beta-
 \sqrt{\left(M^2\sin^2\beta-1\right)^2
 -\dfrac{2(\gamma^2-1)}{\gamma}M^2\sin^2\beta\,Q}}.
 \label{eq:standard-polar-unreduced}
\end{equation}

\citet{pratt1991} expressed the equilibrium relation as a quadratic in
$X_{\mathrm P}=\rho_1/\rho_2$ and reported both density-ratio roots.
To simplify the subsequent equations, define
\begin{equation}
 u=M^2\sin^2\beta,
 \qquad
 q'=\frac{2(\gamma-1)}{\gamma}Q,
 \qquad
 D=(u-1)^2-(\gamma+1)u q'.
 \label{eq:compact-variables}
\end{equation}
With these definitions, \eqref{eq:standard-polar-unreduced} reduces to
\begin{equation}
 \frac{\tan\beta}{\tan(\beta-\theta)}
 =\frac{(\gamma+1)u}{1+\gamma u-\sqrt{D}}.
 \label{eq:standard-polar}
\end{equation}
To retain both density-ratio roots, introduce the reciprocal density
ratio
\begin{equation}
 r=\frac{\rho_1}{\rho_2}
 =\frac{\tan(\beta-\theta)}{\tan\beta}.
 \label{eq:reciprocal-density-ratio}
\end{equation}
The mass, normal-momentum and total-energy jump conditions then give
\begin{equation}
 (\gamma+1)u r^2-2(1+\gamma u)r
 +(\gamma-1)u+2+q'=0.
 \label{eq:rh-quadratic}
\end{equation}
This quadratic contains both density-ratio roots, and its discriminant is
$4D$.

Solving \eqref{eq:rh-quadratic} gives
\begin{equation}
 r_H=\frac{1+\gamma u-\sqrt{D}}{(\gamma+1)u},
 \qquad
 r_L=\frac{1+\gamma u+\sqrt{D}}{(\gamma+1)u}.
 \label{eq:density-roots}
\end{equation}
Since the compression ratio is $\rho_2/\rho_1 = 1/r$, $r_H$
and $r_L$ are the high- and low-compression density-ratio roots,
respectively.

The two density-ratio roots coalesce where $D=0$. In the detonation domain
$u>1$, the relevant zero of $D$ is
\begin{equation}
 u_{\mathrm{CJ}}=M_{\mathrm{CJ}}^{2}
 =\mathcal{A}+\sqrt{\mathcal{A}^{2}-1},
 \qquad
 \mathcal{A}=1+\frac{\gamma^{2}-1}{\gamma}Q,
 \label{eq:u-cj}
\end{equation}
and real density-ratio roots require $u\ge u_{\mathrm{CJ}}$, the
Chapman--Jouguet (CJ) bound on the normal Mach number
\citep{fickett2000,pratt1991}.

For $M>M_{\mathrm{CJ}}$, the bound \eqref{eq:u-cj} restricts oblique waves
to $\beta\ge\beta_{\mathrm{CJ}}=\arcsin(M_{\mathrm{CJ}}/M)$. At
$\beta_{\mathrm{CJ}}$ the two density-ratio roots coalesce at
$r_{\mathrm{CJ}}=(1+\gamma u_{\mathrm{CJ}})/[(\gamma+1)u_{\mathrm{CJ}}]$,
and \eqref{eq:reciprocal-density-ratio} gives the deflection
\begin{equation}
 \theta_{\mathrm{CJ}}=\beta_{\mathrm{CJ}}
 -\arctan\!\left(r_{\mathrm{CJ}}\tan\beta_{\mathrm{CJ}}\right).
 \label{eq:theta-cj}
\end{equation}
At $M=M_{\mathrm{CJ}}$ this endpoint degenerates to the normal CJ wave
($\beta=\pi/2$, $\theta=0$). Equations \eqref{eq:u-cj} and
\eqref{eq:theta-cj} are equivalent to equations (28)--(32) of
\citet{pratt1991}.

\FloatBarrier
\subsection{Exact cubic inversion}
\label{sec:cubic}
The preceding quadratic determines the two density-ratio states for a
prescribed wave angle $\beta$. The inverse problem instead prescribes
$\theta$ and solves for the admissible wave angles $\beta$.
Introduce
\begin{equation}
 t=\tan\beta,\qquad \tau=\tan\theta,\qquad m=M^{2}.
 \label{eq:tangent-variables}
\end{equation}
With $\tan(\beta-\theta)=(t-\tau)/(1+t\tau)$ and
$\sin^{2}\beta=t^{2}/(1+t^{2})$, the density ratio
\eqref{eq:reciprocal-density-ratio} and the normal Mach number become
\begin{equation}
 r=\frac{t-\tau}{t(1+t\tau)},
 \qquad
 u=\frac{mt^{2}}{1+t^{2}}.
 \label{eq:tangent-geometry}
\end{equation}
Attached waves satisfy $0<\theta<\beta<\pi/2$, that is,
$0<\tau<t<\infty$.

Substituting \eqref{eq:tangent-geometry} into the quadratic
\eqref{eq:rh-quadratic} and multiplying by the common denominator
$t(1+t^{2})(1+t\tau)^{2}$ gives
\begin{equation}
 \begin{aligned}
  N(t)={}&(\gamma+1)m\,t(t-\tau)^{2}
  -2\bigl[1+(1+\gamma m)t^{2}\bigr](t-\tau)(1+t\tau)\\
  &+t\bigl\{(2+q')+\bigl[2+q'+(\gamma-1)m\bigr]t^{2}\bigr\}(1+t\tau)^{2}
  =0.
 \end{aligned}
 \label{eq:numerator}
\end{equation}
Expanding $N(t)$ gives the exact factorisation
$N(t)=(1+t^{2})\,C_{3}(t)$. Since
$t(1+t^{2})(1+t\tau)^{2}>0$ in the attached domain,
\eqref{eq:rh-quadratic} holds exactly when
\begin{equation}
 C_{3}(t)=a_{3}t^{3}+a_{2}t^{2}+a_{1}t+a_{0}=0,
 \label{eq:cubic}
\end{equation}
with
\begin{equation}
 \left.
 \begin{aligned}
  a_{3}&=\tau^{2}\bigl[q'+(\gamma-1)m+2\bigr],\\
  a_{2}&=2\tau\bigl[q'-(m-1)\bigr],\\
  a_{1}&=q'+\tau^{2}\bigl[(\gamma+1)m+2\bigr],\\
  a_{0}&=2\tau.
 \end{aligned}
 \right\}
 \label{eq:cubic-coefficients}
\end{equation}
The heat release enters only through the additive term in
\begin{equation}
 C_{3}(t)=C_{3}^{(0)}(t)+q'\,t(1+\tau t)^{2},
 \label{eq:additive}
\end{equation}
where $C_{3}^{(0)}=C_{3}\big|_{q'=0}$. For $Q=0$ the additive term
vanishes, the discontinuity is an oblique shock, and \eqref{eq:cubic}
becomes (after division by $2\tau$) the classical oblique-shock
$\theta$--$\beta$--$M$ cubic in $\tan\beta$
\citep{wellmann1972,emanuel2001},
\begin{equation}
 \bigl[1+\tfrac{1}{2}(\gamma-1)m\bigr]\tau t^{3}
 +(1-m)t^{2}
 +\bigl[1+\tfrac{1}{2}(\gamma+1)m\bigr]\tau t
 +1=0.
 \label{eq:inert-cubic}
\end{equation}

\subsection{Branch-complete physical selection}
\label{sec:selection}

For given $(\theta,M,\gamma,Q)$, let $t_{1}$, $t_{2}$ and $t_{3}$ denote
the three algebraic roots of \eqref{eq:cubic}, counted with multiplicity.
Of the three Vieta relations for a cubic, only the product relation is
needed here:
\begin{equation}
  t_{1}t_{2}t_{3}=-\frac{a_{0}}{a_{3}}
  =-\frac{2}{\tau\,[q'+(\gamma-1)m+2]}<0.
 \label{eq:cubic-vieta}
\end{equation}
The right-hand side is negative because $\tau>0$, $\gamma>1$, $m>0$
and $q'\geq0$. It shows that the three roots cannot all be positive.
Thus at most two roots can satisfy the attachment condition $t>\tau>0$.
Admissible high-compression wave-angle solutions are selected by an ordered
filter: a candidate must be (i)~real;
(ii)~attached, $t>\tau$; (iii)~above the CJ bound,
$u\ge u_{\mathrm{CJ}}$; and (iv)~on the high-compression branch.

The elimination leading to \eqref{eq:cubic} removes the density ratio, so an
algebraic root no longer identifies the density-ratio branch from which it
originated. Step (iv) restores this information. Rearranging
\eqref{eq:density-roots} gives
\begin{equation*}
 1+\gamma u-(\gamma+1)u r_H=+\sqrt{D},
 \qquad
 1+\gamma u-(\gamma+1)u r_L=-\sqrt{D}.
\end{equation*}
For a candidate root $t$, evaluate $u$ and $r$ from
\eqref{eq:tangent-geometry} and define the corresponding signed branch
quantity by
\begin{equation}
 \sigma=1+\gamma u-(\gamma+1)ur.
 \label{eq:branch-sign}
\end{equation}
Thus $\sigma>0$ identifies the high-compression branch, $\sigma<0$
identifies the low-compression branch, and $\sigma=0$ occurs where the two
density-ratio roots coalesce. The discriminant cannot distinguish the
branches: squaring either branch relation gives $D=\sigma^{2}$, so $D\ge0$
holds on both branches and contains no sign information.

For $\theta_{\mathrm{CJ}}<\theta<\theta_{\max}$, two distinct wave-angle
roots pass the filter: the smaller gives the weak overdriven solution and the
larger the strong solution. The roots coalesce at the maximum deflection
$\theta_{\max}$, discussed in the next subsection. The stated interval is
established at the end of \S\ref{sec:detachment}.

The classical oblique-shock cubic \eqref{eq:inert-cubic} is recovered from
\eqref{eq:cubic} when $Q=0$. We solve the reactive cubic by Cardano's method.
Write \eqref{eq:cubic} as $at^{3}+bt^{2}+ct+d=0$ with
$(a,b,c,d)=(a_{3},a_{2},a_{1},a_{0})$ from \eqref{eq:cubic-coefficients};
in the attached domain $a>0$. The substitution $t=z-b/(3a)$ removes
the quadratic term and gives the depressed cubic
\begin{equation}
 z^{3}+pz+q=0,
 \label{eq:depressed}
\end{equation}
with
\begin{equation}
 p=\frac{3ac-b^{2}}{3a^{2}},
 \qquad
 q=\frac{27a^{2}d-9abc+2b^{3}}{27a^{3}},
 \label{eq:pq}
\end{equation}
and the discriminant quantity
\begin{equation}
 \Delta_{C}=\Bigl(\frac{q}{2}\Bigr)^{2}+\Bigl(\frac{p}{3}\Bigr)^{3}.
 \label{eq:cardano-disc}
\end{equation}
The sign of $\Delta_{C}$ separates the three root patterns.

For $\Delta_{C}<0$ (hence $p<0$) all three algebraic roots are real
and distinct. Cardano's radicals then pass through complex
intermediates even though every root is real, so we use the
equivalent trigonometric form, which keeps the arithmetic real
\citep{nickalls1993}:
\begin{equation}
 t_{k}=-\frac{b}{3a}+2\sqrt{-\frac{p}{3}}
 \cos\!\left(\phi-\frac{2\pi k}{3}\right),
 \qquad k=0,1,2,
 \label{eq:trig-roots}
\end{equation}
where
\begin{equation}
 \phi=\frac{1}{3}\arccos\!\left(\frac{3q}{2p}\sqrt{-\frac{3}{p}}\right).
 \label{eq:trig-angle}
\end{equation}
The coexisting weak and strong solutions fall in this case: with two
distinct positive roots, the third root is negative by the product
relation in \eqref{eq:cubic-vieta}, so all three roots are real and
distinct. For
$\theta<\theta_{\mathrm{CJ}}$ the two positive roots instead correspond
to the strong high-compression and low-compression candidates
(appendix~\ref{app:polar-geometry}).

When $\Delta_{C}=0$, the cubic has repeated roots. If
$(p,q)\ne(0,0)$, let $w=\sqrt[3]{-q/2}$. The roots are the simple root
$t_{s}=-b/(3a)+2w$ and the double root $t_{d}=-b/(3a)-w$. At
detachment, the weak and strong attached roots merge into the
positive double root $t_{d}$. The product relation in
\eqref{eq:cubic-vieta} gives $t_{s}t_{d}^{2}=-a_{0}/a_{3}<0$; hence the
remaining root satisfies $t_{s}<0$ and lies outside the attached-wave
domain $t>\tau>0$. If
$p=q=0$, all three algebraic roots equal $-b/(3a)$. For the present
coefficients, \eqref{eq:cubic-vieta} gives
$t_{*}^{3}=-d/a=-a_{0}/a_{3}<0$; the triple root is therefore negative
and does not represent an attached wave.

For $\Delta_{C}>0$ the single real algebraic root is
\begin{equation}
 t=-\frac{b}{3a}
 +\sqrt[3]{-\frac{q}{2}+\sqrt{\Delta_{C}}}
 +\sqrt[3]{-\frac{q}{2}-\sqrt{\Delta_{C}}},
 \label{eq:cardano-root}
\end{equation}
where both cube roots are taken real. If the non-real conjugate pair is
$z,\overline{z}$ and the real root is $t_{r}$, then
$z\overline{z}=|z|^{2}>0$ and \eqref{eq:cubic-vieta} gives
$t_{r}|z|^{2}=-a_{0}/a_{3}<0$. Thus $t_{r}<0$, so no attached candidate
exists: the deflection lies beyond the maximum of the attached polar.

\begin{figure}[!ht]
 \centering
 \includegraphics[width=\linewidth]{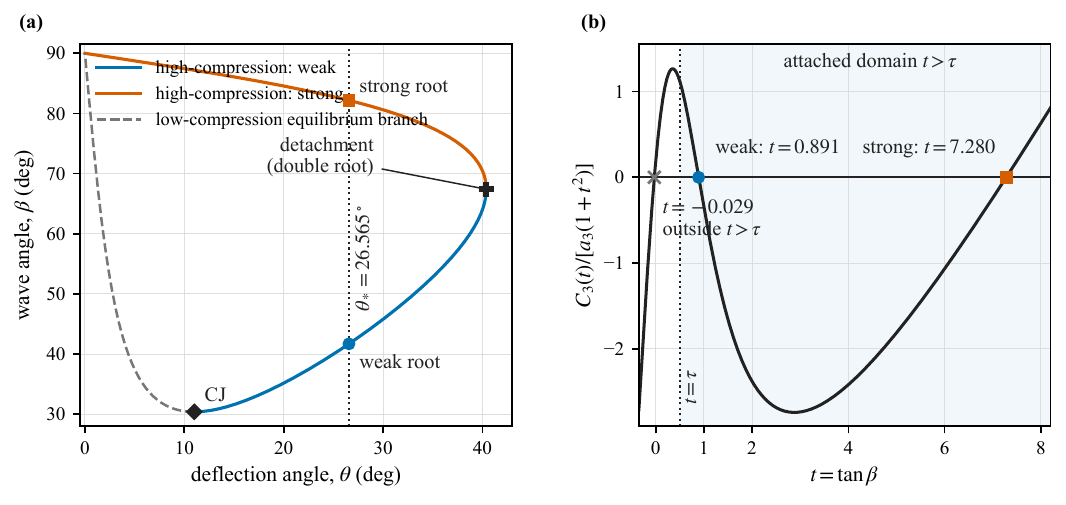}
 \caption{Three distinct real roots for $\Delta_{C}<0$ at
 $\gamma=1.3$, $M=7$, $Q=10$ and
 $\theta=26.565051^{\circ}$. (a) The prescribed-deflection line
 intersects the weak and strong high-compression branches. (b) The
 scaled cubic has three real zeros, of which the two positive zeros lie
 in the attached domain. The markers are common to both panels: circle,
 weak root; square, strong root; filled plus, double root at detachment;
 diamond, CJ endpoint; cross, root outside the attached domain
 $t>\tau$.}
 \label{fig:cardano-negative}
\end{figure}

\begin{figure}[!ht]
 \centering
 \includegraphics[width=\linewidth]{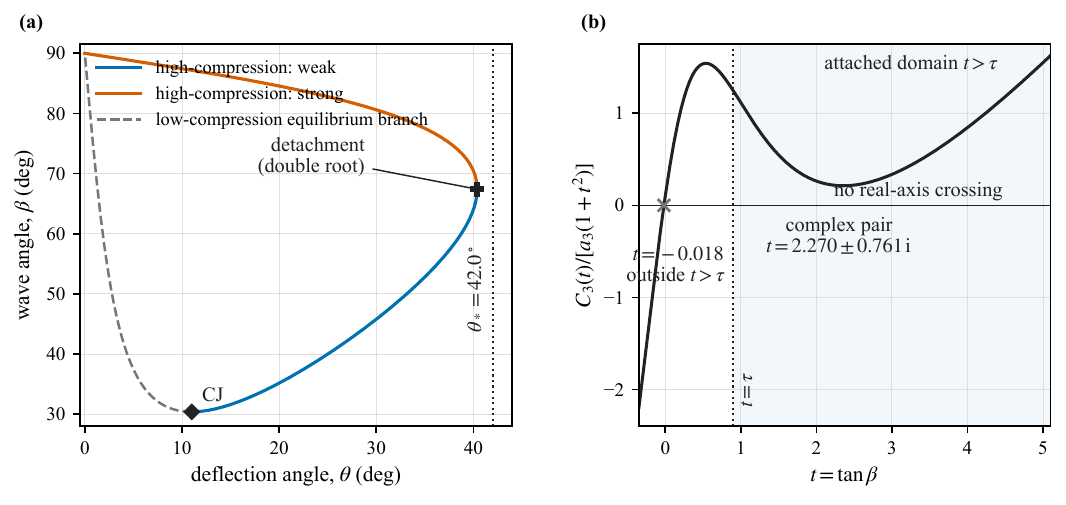}
 \caption{One-real-root case for $\Delta_{C}>0$ at
 $\gamma=1.3$, $M=7$, $Q=10$ and $\theta=42^{\circ}$. (a) The
 prescribed deflection exceeds the maximum of the attached polar. (b)
 The only real zero of the scaled cubic is negative, and the attached
 domain contains no root. Markers as in
 figure~\ref{fig:cardano-negative}.}
 \label{fig:cardano-positive}
\end{figure}

To illustrate the three discriminant cases on the same equilibrium polar,
we fix
\begin{equation*}
 \gamma=1.3,\qquad M=7,\qquad Q=10,
\end{equation*}
and vary only the prescribed deflection $\theta$.
Figures~\ref{fig:cardano-negative}--%
\ref{fig:cardano-positive} use a common layout. Panel (a) shows the polar
in the $(\theta,\beta)$ plane and the prescribed-deflection line. Panel
(b) plots the scaled cubic $C_{3}(t)/[a_{3}(1+t^{2})]$ against real $t$,
with the attached domain $t>\tau$ shaded. Since $a_{3}>0$ and
$1+t^{2}>0$, the scaling preserves the real roots and their
multiplicities.

For $\Delta_{C}<0$, a representative prescribed deflection and the
resulting roots are
\begin{equation*}
 \begin{gathered}
  \theta=\arctan(1/2)=26.565051^{\circ},\qquad
  \tau=0.5,\qquad \Delta_{C}=-17.073613,\\
  t\in\{-0.028943,\ 0.890644,\ 7.279764\},\\
  \beta_{\mathrm{weak}}=41.689675^{\circ},\qquad
  \beta_{\mathrm{strong}}=82.178397^{\circ}.
 \end{gathered}
\end{equation*}
The two positive roots correspond to the weak and strong high-compression
attached solutions, while the negative root lies outside $t>\tau$.

\FloatBarrier

For $\Delta_{C}=0$, the repeated-root data at the maximum deflection are
\begin{equation*}
 \begin{gathered}
  \theta=\theta_{\max}=40.382949^{\circ},\qquad
  \tau=0.850554,\qquad \Delta_{C}=0,\\
  t_{s}=-0.019111,\qquad
  t_{d}=2.402545,\qquad
  \beta_{d}=67.401690^{\circ}.
 \end{gathered}
\end{equation*}
The simple root is negative, while the weak and strong roots merge at the
positive double root $t_{d}$.

\begin{figure}[!ht]
 \centering
 \includegraphics[width=\linewidth]{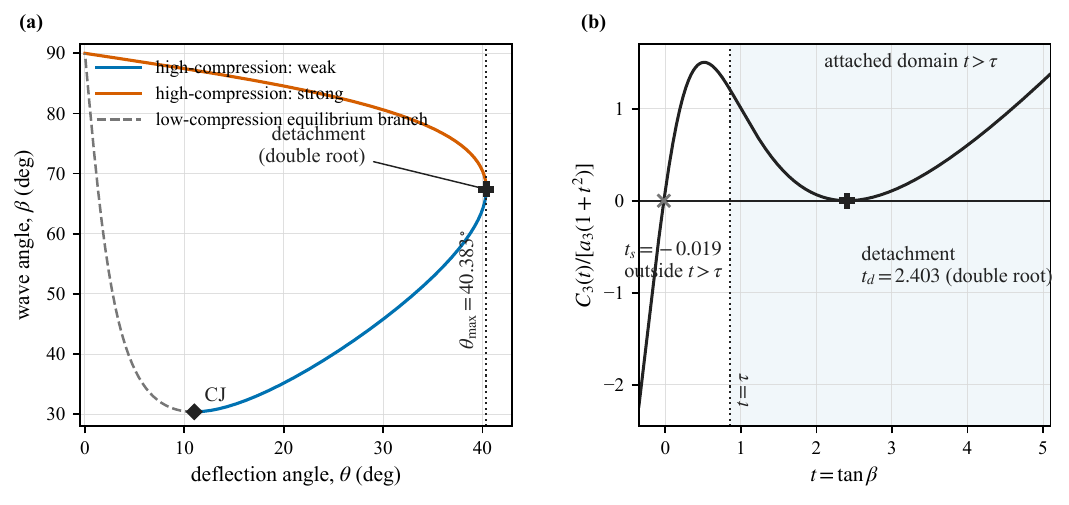}
 \caption{Repeated-root case for $\Delta_{C}=0$ at
 $\gamma=1.3$, $M=7$, $Q=10$ and
 $\theta=\theta_{\max}=40.382949^{\circ}$. (a) The weak and strong
 branches meet at the maximum-deflection point. (b) The scaled cubic is
 tangent to the real axis at the positive double root; the remaining
 simple root is negative. Markers as in
 figure~\ref{fig:cardano-negative}.}
 \label{fig:cardano-zero}
\end{figure}
\FloatBarrier

For $\Delta_{C}>0$, a representative prescribed deflection above the
maximum gives
\begin{equation*}
 \begin{gathered}
  \theta=42^{\circ}>\theta_{\max},\qquad
  \tau=0.900404,\qquad \Delta_{C}=0.725583,\\
  t\in\{-0.018184,\ 2.269595\pm0.761346\,\mathrm{i}\}.
 \end{gathered}
\end{equation*}
The only real root is negative. Thus no real root lies in the attached
domain, consistently with the absence of an intersection between the
prescribed-deflection line and the attached polar.

\FloatBarrier

The wave angle associated with a retained real root is
$\beta=\arctan t$. Figures~\ref{fig:cardano-negative}--%
\ref{fig:cardano-positive} also show why an algebraic root index cannot
be assigned to a fixed physical branch across the parameter domain. Each
real root must therefore be classified by the filter (i)--(iv).

\subsection{Exact fixed-$Q$ detachment point}
\label{sec:detachment}

At detachment, the weak and strong solutions merge and the cubic
\eqref{eq:cubic} has a double wave-angle root ($\Delta_{C}=0$).
Section~\ref{sec:selection} used this condition to classify solutions at a prescribed
deflection; we now use it to derive explicit detachment formulae. For
compactness, define
\begin{equation}
 A=q'+(\gamma-1)m+2,\qquad B=q'-(m-1),\qquad C=(\gamma+1)m+2.
 \label{eq:ABC}
\end{equation}
These quantities depend only on $M$, $\gamma$ and $Q$. With these
definitions, the cubic \eqref{eq:cubic} can be written directly in
$(\tau,t)$ as
\begin{equation}
 A\tau^{2}t^{3}+2B\tau t^{2}+(q'+C\tau^{2})t+2\tau=0.
 \label{eq:cubic-tau-t}
\end{equation}
To simplify the detachment calculation, set
\begin{equation*}
 y=\tau t,\qquad s=\tau^{2}.
\end{equation*}
Multiplying \eqref{eq:cubic-tau-t} by $\tau$ then gives
\begin{equation}
 G(y,s)=Ay^{3}+2By^{2}+(q'+Cs)y+2s=0.
 \label{eq:G-ys}
\end{equation}

For fixed $s>0$, the map $y=\tau t$ is linear and preserves root
multiplicity. Hence the double root $t_{d}$ at detachment corresponds to a
double root $y_{d}$ of $G(\mathord{\cdot},s_{d})$ and therefore satisfies
\begin{equation*}
 \begin{gathered}
  G(y_{d},s_{d})=0,\qquad G_{y}(y_{d},s_{d})=0,\\
  G_{y}\equiv\frac{\partial G}{\partial y}
  =3Ay^{2}+4By+q'+Cs.
 \end{gathered}
\end{equation*}

Combining the two double-root conditions gives
\begin{equation*}
 G-yG_{y}=-2Ay^{3}-2By^{2}+2s=0
\end{equation*}
and hence the recovery relation
\begin{equation}
 s=y^{2}(Ay+B).
 \label{eq:s-recovery}
\end{equation}
Substitution into $G_{y}=0$ then eliminates $s$:
\begin{equation}
 AC\,y_{d}^{3}+(BC+3A)\,y_{d}^{2}+4B\,y_{d}+q'=0.
 \label{eq:detachment-cubic}
\end{equation}
The fixed-$Q$ detachment condition therefore reduces to a cubic in the
double root $y_{d}$. It can be solved by the same reduction to a depressed
cubic and the same discriminant-based formulae used for the wave-angle
cubic \eqref{eq:cubic}; only the unknown and coefficients differ.

Each real root $y_{d}$ of \eqref{eq:detachment-cubic} yields a candidate
boundary point. The remaining quantities are recovered as
\begin{equation*}
 s_{d}=y_{d}^{2}(Ay_{d}+B),\qquad
 \tau_{d}=\sqrt{s_{d}},\qquad
 t_{d}=\frac{y_{d}}{\sqrt{s_{d}}}.
\end{equation*}
The corresponding angles are
\begin{equation}
 \theta_{\max}=\arctan\sqrt{s_{d}},
 \qquad
 \beta_{d}=\beta_{\max}
 =\arctan\bigl(y_{d}/\sqrt{s_{d}}\bigr).
 \label{eq:detachment-angles}
\end{equation}
Here $u_{d}=mt_{d}^{2}/(1+t_{d}^{2})$. A candidate is retained only if
\begin{equation*}
 \begin{gathered}
  y_{d}>0,\qquad s_{d}>0,\qquad
  t_{d}=\frac{y_{d}}{\sqrt{s_{d}}}>\sqrt{s_{d}},\\
  u_{d}\ge u_{\mathrm{CJ}},\qquad \sigma\ge0.
 \end{gathered}
\end{equation*}
These conditions, with $\sigma$ evaluated at $t_{d}$, restate the filter
(i)--(iv) at the double root. If several candidates pass, the detachment point is the one with
the largest $s_{d}$, because $\theta=\arctan\sqrt{s}$ increases with $s$.

An explicit maximum-deflection relation was derived by
\citet{townend1970} with the wave-type parameter $F$ held
fixed. Since $F$ varies along a fixed-$Q$ polar, that condition is distinct
from the fixed-$Q$ double-root condition derived here.

As a consistency check, the discriminant of $C_{3}$ with respect to $t$ is
quartic in $s$ and contains the trivial factor $s$. Removing this factor
leaves a cubic condition, in agreement with
\eqref{eq:detachment-cubic}. For $q'=0$, the detachment cubic reduces to
\begin{equation*}
 y_{d}\bigl[ACy_{d}^{2}+(BC+3A)y_{d}+4B\bigr]=0.
\end{equation*}
The root $y_{d}=0$ is inadmissible because a physical candidate requires
$y_{d}>0$. The remaining factor gives
\begin{equation}
 ACy_{d}^{2}+(BC+3A)y_{d}+4B=0,
 \label{eq:inert-detachment}
\end{equation}
where $A=(\gamma-1)m+2$, $B=1-m$ and $C=(\gamma+1)m+2$.
Equation~\eqref{eq:s-recovery} gives $t_{d}^{2}=1/(Ay_{d}+B)$, so the
change of variable
\begin{equation*}
 x=\sin^{2}\beta_{d}=\frac{1}{Ay_{d}+B+1}
\end{equation*}
transforms \eqref{eq:inert-detachment} into
\begin{equation}
 2\gamma M^{4}x^{2}
 -\bigl[(\gamma+1)M^{4}-4M^{2}\bigr]x
 -\bigl[(\gamma+1)M^{2}+2\bigr]=0,
 \label{eq:standard-shock-detachment}
\end{equation}
the standard oblique-shock detachment quadratic \citep{wellmann1972}.

The double-root structure also fixes the geometry of the attached
polar; the proofs, which combine the root count of
\S\ref{sec:selection} with continuity along the branches, are given
in appendix~\ref{app:polar-geometry}. First, the double root cannot
occur at the CJ bound: $u_{d}>u_{\mathrm{CJ}}$, so
$\beta_{d}>\beta_{\mathrm{CJ}}$ and the high-compression branch
splits into a weak segment $\beta\in[\beta_{\mathrm{CJ}},\beta_{d}]$
and a strong segment $\beta\in[\beta_{d},\pi/2]$. Second, the
deflection attains its strict maximum over all attached equilibrium
states at the double root; in particular
$\theta_{\mathrm{CJ}}<\theta_{\max}$. Third, along the weak segment
the deflection increases strictly with the wave angle from
$\theta_{\mathrm{CJ}}$ to $\theta_{\max}$; the weak overdriven
solution therefore exists precisely on
$\theta_{\mathrm{CJ}}<\theta<\theta_{\max}$, as stated in
\S\ref{sec:selection}. Finally, beyond the CJ point the
low-compression branch has $\theta<\theta_{\mathrm{CJ}}$, so for
$\theta<\theta_{\mathrm{CJ}}$ the two positive roots of the cubic are
the strong high-compression and low-compression solutions.

\section{Exact structure of the solution space}
\label{sec:structure}

\begin{figure}[!ht]
 \centering
 \includegraphics[width=\linewidth]{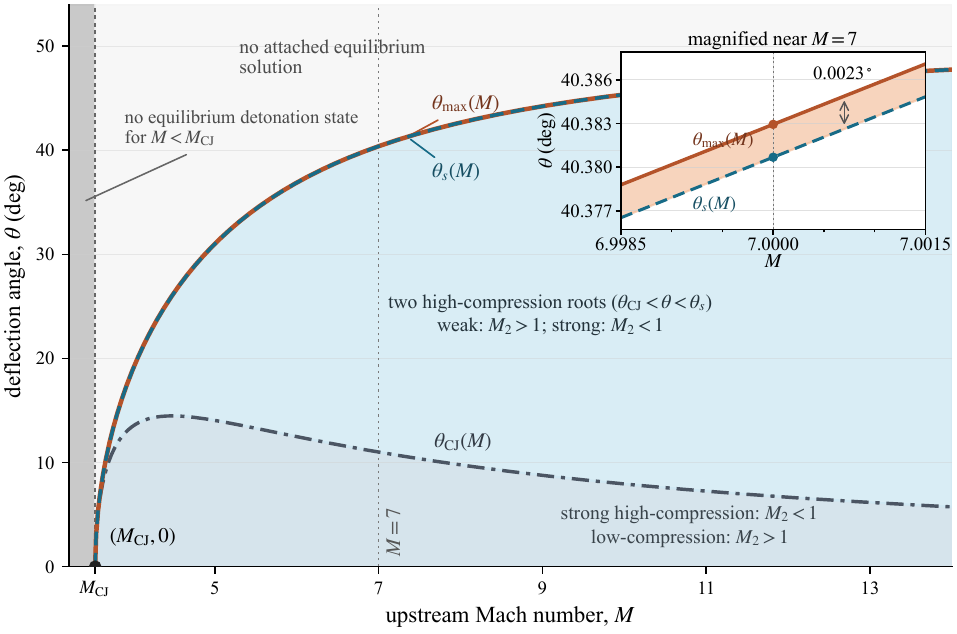}
 \caption{Morphology of the solution domain in the $(M,\theta)$ plane for
 $\gamma=1.3$ and $Q=10$, for which $M_{\mathrm{CJ}}=3.5406$. The domain is
 partitioned by three closed-form curves: the CJ deflection
 $\theta_{\mathrm{CJ}}(M)$ from \eqref{eq:u-cj} and \eqref{eq:theta-cj}, the
 detachment boundary $\theta_{\max}(M)$ from \eqref{eq:detachment-cubic} and
 \eqref{eq:detachment-angles}, and the sonic line $\theta_{s}(M)$ derived in
 \S\ref{sec:sonic}. The sonic line subdivides the weak high-compression
 branch by the downstream Mach number $M_{2}$; it is not the weak--strong
 boundary. It runs so close to the detachment boundary that the two are
 graphically indistinguishable in the main panel. The inset resolves them
 near $M=7$, where they bound a strip of width $0.0023^{\circ}$. Inside this
 strip the weak high-compression solution is subsonic downstream; below it
 the same solution is supersonic. The downstream Mach numbers annotated in
 each region are established in \S\ref{sec:classification}.}
 \label{fig:morphology}
\end{figure}

This section establishes the global structure of the equilibrium
solution space. Section~\ref{sec:sonic} derives the downstream sonic
locus in closed form. Section~\ref{sec:classification} locates this
locus on the physical polar and classifies the solution branches by
downstream Mach number.
Figure~\ref{fig:morphology} provides an overview of this structure.

Figure~\ref{fig:morphology} maps the solution domain in the
$(M,\theta)$ plane for fixed $\gamma$ and $Q$. The domain is partitioned
by three closed-form curves: the CJ deflection
$\theta_{\mathrm{CJ}}(M)$ of \eqref{eq:theta-cj}, the detachment
boundary $\theta_{\max}(M)$ of \eqref{eq:detachment-angles}, and the
sonic line $\theta_{s}(M)$ derived below. Within the present equilibrium
model, no real equilibrium detonation state exists for
$M<M_{\mathrm{CJ}}$. For
$M>M_{\mathrm{CJ}}$, the strong high-compression and low-compression
roots coexist when
$0<\theta<\theta_{\mathrm{CJ}}(M)$, whereas the weak and strong
high-compression roots coexist when
$\theta_{\mathrm{CJ}}(M)<\theta<\theta_{\max}(M)$. No attached
equilibrium solution exists above $\theta_{\max}(M)$. The sonic line
lies between $\theta_{\mathrm{CJ}}(M)$ and $\theta_{\max}(M)$, close
to the detachment boundary. We show below that
the weak high-compression solution is downstream supersonic for
$\theta_{\mathrm{CJ}}(M)<\theta<\theta_{s}(M)$. In the narrow interval
$\theta_{s}(M)<\theta<\theta_{\max}(M)$, the same solution is downstream
subsonic.

\FloatBarrier

\subsection{Exact closed-form sonic locus}
\label{sec:sonic}

The cubic inversion of \S\ref{sec:inversion} determines admissible wave
angles for a prescribed deflection but does not classify the associated
downstream flow as supersonic or subsonic. We therefore derive the downstream
total-sonic locus, defined by $M_{2}=1$. It is distinct from the detachment
locus, where the weak and strong roots coalesce, and from the normal-sonic CJ
condition. Earlier analyses showed qualitatively that a total-sonic point lies
on the weak high-compression branch close to detachment
\citep{powersgonthier1992,emanuel2004}. Here the sonic condition reduces to a
quadratic equation in $u$, yielding the locus explicitly.

To express $M_{2}$ in terms of $u$ and $r$, we note that the tangential
velocity is unchanged across the discontinuity, while mass conservation
reduces the normal velocity by the factor $r$. The normal-momentum jump
then gives the pressure ratio
\begin{equation}
 P=\frac{p_{2}}{p_{1}}=1+\gamma u(1-r).
 \label{eq:pressure-ratio}
\end{equation}
Both real density-ratio roots in \eqref{eq:density-roots} satisfy
$0<r<1$ by \eqref{eq:app-density-bounds}, so $P>1$. The ideal-gas law
then gives $T_{2}/T_{1}=rP$.
With the velocity components normalised by the upstream sound speed, the
squared upstream normal and tangential components are $u$ and $m-u$,
respectively. The downstream Mach number is therefore
\begin{equation}
 M_{2}^{2}=\frac{u r^{2}+(m-u)}{rP}
 =M_{n2}^{2}+\frac{m-u}{rP},
 \qquad
 M_{n2}^{2}=\frac{u r}{P}.
 \label{eq:downstream-mach}
\end{equation}
Because $rP>0$, the sonic condition $M_{2}=1$ is equivalent to the
quadratic
\begin{equation}
 \mathcal{S}(r)\equiv(\gamma+1)u\,r^{2}-(1+\gamma u)\,r+(m-u)=0,
 \label{eq:sonic-quadratic}
\end{equation}
Indeed, $M_{2}^{2}-1=\mathcal{S}(r)/(rP)$. Hence $\mathcal{S}>0$ and
$\mathcal{S}<0$ correspond to downstream supersonic and subsonic states,
respectively.

At a sonic equilibrium state, $r$ satisfies \eqref{eq:rh-quadratic} and
\eqref{eq:sonic-quadratic} simultaneously. The two equations have the
same coefficient of $r^{2}$. Subtracting them therefore gives the sonic
density ratio directly:
\begin{equation}
 r_{*}=\frac{\gamma u+q'+2-m}{1+\gamma u}.
 \label{eq:sonic-density-ratio}
\end{equation}
Substituting \eqref{eq:sonic-density-ratio} into \eqref{eq:rh-quadratic}
eliminates $r$. Multiplying by $(1+\gamma u)^{2}$ clears the denominator.
The cubic terms in $u$ cancel identically, leaving the quadratic
\begin{equation}
 E(u)=c_{2}u^{2}+c_{1}u+c_{0}=0,
 \label{eq:E-quadratic}
\end{equation}
with
\begin{equation}
 \left.
 \begin{aligned}
  c_{2}&=\gamma\bigl[(\gamma+2)q'-2(m-1)\bigr],\\
  c_{1}&=(\gamma+1)(m-1-q')^{2}
  +2\bigl[(\gamma-1)(m-1)+q'\bigr],\\
  c_{0}&=2(m-1)-q',
 \end{aligned}
 \right\}
 \label{eq:E-coefficients}
\end{equation}

For $\gamma>1$, $Q>0$ and $M>M_{\mathrm{CJ}}$, these coefficients have
the fixed signs
\begin{equation*}
 c_{2}<0,\qquad c_{1}>0,\qquad c_{0}>0,
\end{equation*}
as proved in appendix~\ref{app:sonic-coefficient-signs}. Hence
$c_{1}^{2}-4c_{2}c_{0}>c_{1}^{2}$: in the conventional quadratic
formula, $(-c_{1}-\sqrt{c_{1}^{2}-4c_{2}c_{0}})/(2c_{2})$ is positive,
whereas the companion root is negative. Thus $E$ has exactly one positive
root. The unique positive sonic candidate is therefore
\begin{equation}
 u_{s}=\frac{c_{1}+\sqrt{c_{1}^{2}-4c_{2}c_{0}}}{-2c_{2}},
 \qquad
 \beta_{s}=\arcsin\sqrt{u_{s}/m},
 \label{eq:sonic-root}
\end{equation}
With $r_{*}$ evaluated at $u_{s}$, the deflection angle is
\begin{equation}
\tan\theta_{s}
 =\frac{\tan\beta_{s}(1-r_{*})}
 {1+r_{*}\tan^{2}\beta_{s}}.
 \label{eq:sonic-deflection}
\end{equation}
For fixed $(\gamma,Q)$, varying $M$ in
\eqref{eq:sonic-root}--\eqref{eq:sonic-deflection} traces the sonic
line $\theta_{s}(M)$ shown in figure~\ref{fig:morphology}; its physical
admissibility and branch location are established next.

\subsection{Physical location and branch classification}
\label{sec:classification}

Section~\ref{sec:sonic} has shown that $E$ possesses exactly one positive
root $u_{s}$. We now locate this root on the physical polar. Direct
substitution gives the endpoint values
\begin{equation}
 \begin{aligned}
  E(u_{\mathrm{CJ}})&=(\gamma+1)\,u_{\mathrm{CJ}}
  \left(m-u_{\mathrm{CJ}}\right)^{2}>0,\\
  E(m)&=-\bigl[(\gamma-1)m+q'+2\bigr]
  \left(m^{2}-2\mathcal{A}m+1\right)<0,
 \end{aligned}
 \label{eq:E-endpoints}
\end{equation}
The factor $m^{2}-2\mathcal{A}m+1$ is positive because
$m>u_{\mathrm{CJ}}$, where $u_{\mathrm{CJ}}$ is the larger root of
$m^{2}-2\mathcal{A}m+1=0$. Thus, $E$ changes sign between
$u_{\mathrm{CJ}}$ and $m$. By continuity, it has a root in this interval.
Since $E$ has exactly one positive root, this root is $u_{s}$. To restore
the branch information lost when $r$ was eliminated, first note that the
low-compression root satisfies
\begin{equation}
 \mathcal{S}(r_{L})=r_{L}\sqrt{D}+(m-u)>0
 \qquad\text{for }u_{\mathrm{CJ}}<u<m.
 \label{eq:S-low-compression}
\end{equation}
At $u=u_{s}$, the ratio $r_{*}$ satisfies both
\eqref{eq:rh-quadratic} and \eqref{eq:sonic-quadratic}, so it must equal
either $r_{H}$ or $r_{L}$. Equation~\eqref{eq:S-low-compression}
excludes the latter, and hence $r_{*}(u_{s})=r_{H}(u_{s})$. Moreover,
$u=m\sin^{2}\beta$ is one-to-one
over the attached wave-angle range. The unique $u_{s}$ therefore
determines a unique sonic state on the high-compression polar.

It remains to locate this state relative to detachment. Let
$(y_{d},s_{d})$ denote the detachment pair selected by the complete
physical admissibility conditions of \S\ref{sec:detachment}, and let
$\mathcal{S}_{d}$ denote the value of $\mathcal{S}$ at this state. Since
\begin{equation*}
 M_{2}^{2}-1=\frac{\mathcal{S}}{rP}
\end{equation*}
and $rP>0$, $\mathcal{S}$ has the same sign as $M_{2}^{2}-1$. We now use
the detachment variables $y=\tau t$ and $s=\tau^{2}$ introduced in
\S\ref{sec:detachment}. They give
\begin{equation*}
 u=\frac{my^{2}}{s+y^{2}},
 \qquad
 r=\frac{y-s}{y(1+y)}.
\end{equation*}
Substitution into $\mathcal{S}$ yields
\begin{equation}
 \mathcal{S}=-\frac{myL+(y+1)(y-s)}{y(y+1)^{2}},
 \qquad
 L=\gamma(y-s)-s-1.
 \label{eq:S-ys}
\end{equation}
The cubic $G$ and its derivative with respect to $y$ satisfy the identity
\begin{equation}
 y(y+1)G_{y}-(3y+1)G
 =2\bigl[my^{2}L+(y+1)\bigl(y^{2}-2sy-s\bigr)\bigr],
 \label{eq:G-combination}
\end{equation}
At the double root, $G=G_{y}=0$. The identity therefore gives
\begin{equation*}
 my^{2}L=-(y+1)(y^{2}-2sy-s).
\end{equation*}
Consequently,
\begin{equation*}
myL=-\frac{y+1}{y}(y^2-2sy-s).
\end{equation*}
Substituting this relation into the numerator of \eqref{eq:S-ys} yields
\begin{equation*}
 \begin{aligned}
  myL+(y+1)(y-s)
  &=\frac{y+1}{y}\bigl[y(y-s)-\bigl(y^{2}-2sy-s\bigr)\bigr]\\
  &=\frac{s(y+1)^{2}}{y}.
 \end{aligned}
\end{equation*}
Dividing the simplified numerator by $-y(y+1)^{2}$, as prescribed by
\eqref{eq:S-ys}, gives
\begin{equation}
 \mathcal{S}_{d}
 =-\frac{s_{d}}{y_{d}^{2}}
 =-\frac{1}{t_{d}^{2}}
 =-\cot^{2}\beta_{d}<0.
 \label{eq:detachment-sonic-identity}
\end{equation}
At a non-degenerate detachment point, $\theta_{\max}>0$, so
$s_{d}=\tan^{2}\theta_{\max}>0$; admissibility also requires $y_{d}>0$.
These conditions establish the strict inequality in
\eqref{eq:detachment-sonic-identity}. Let $P_{d}$ and $r_{d}$ denote the
pressure and density ratios at detachment. Since $P_{d}>1$ and $r_{d}>0$,
the Mach-number identity gives

\begin{equation*}
 M_{2,d}^{2}-1=\frac{\mathcal{S}_{d}}{r_{d}P_{d}}<0.
\end{equation*}
Thus, $M_{2,d}^{2}<1$, and hence $M_{2,d}<1$.

Using the definition of $\mathcal{S}$ in
\eqref{eq:sonic-quadratic} together with the high-compression root in
\eqref{eq:density-roots}, we obtain
\begin{equation}
 \begin{aligned}
 \mathcal{S}(r_{H}(u))
 &=r_{H}(u)\bigl[(\gamma+1)u r_{H}(u)-(1+\gamma u)\bigr]+m-u\\
 &=m-u-r_{H}(u)\sqrt{D(u)}.
 \end{aligned}
 \label{eq:S-high-compression}
\end{equation}
Both $r_{H}(u)$ and $\sqrt{D(u)}$ are continuous for
$u\geq u_{\mathrm{CJ}}$, so $\mathcal{S}(r_H(u))$ is continuous on
$[u_{\mathrm{CJ}},u_{d}]$. At the CJ endpoint, $D=0$ and hence
\begin{equation*}
 \mathcal{S}(r_{\mathrm{CJ}})=m-u_{\mathrm{CJ}}>0,
\end{equation*}
whereas \eqref{eq:detachment-sonic-identity} gives
$\mathcal{S}(r_{H}(u_{d}))=\mathcal{S}_{d}<0$. The intermediate value
theorem therefore guarantees a sonic state in
$(u_{\mathrm{CJ}},u_{d})$. The high-compression polar has exactly one
sonic state, already identified as $u_{s}$, so
$u_{\mathrm{CJ}}<u_{s}<u_{d}<m$.

Since $\beta=\arcsin\sqrt{u/m}$ increases with $u$ at fixed $M$, the
ordering in $u$ gives $\beta_{\mathrm{CJ}}<\beta_{s}<\beta_{d}$: the
sonic state lies in the interior of the weak segment. Along this
segment the deflection increases strictly with the wave angle
(\S\ref{sec:detachment}), so
$\theta_{\mathrm{CJ}}<\theta_{s}<\theta_{\max}$.
The complete ordering is therefore
\begin{equation*}
 u_{\mathrm{CJ}}<u_{s}<u_{d}<m,
 \qquad
 \beta_{\mathrm{CJ}}<\beta_{s}<\beta_{d},
 \qquad
 \theta_{\mathrm{CJ}}<\theta_{s}<\theta_{\max}.
\end{equation*}

The downstream Mach number along the high-compression polar follows
from the sign of $\mathcal{S}(r_{H})$. On $[u_{\mathrm{CJ}},m]$ this
quantity is continuous and vanishes only at $u_{s}$, so its sign is
constant on $[u_{\mathrm{CJ}},u_{s})$ and on $(u_{s},m]$. Since
$\mathcal{S}(r_{\mathrm{CJ}})>0$ and $\mathcal{S}_{d}<0$, the weak
branch is supersonic downstream before the sonic point and subsonic
after it, and the entire strong branch $u\in(u_{d},m]$ is subsonic
downstream.

The low-compression branch is supersonic downstream throughout: in
$\mathcal{S}(r_{L})=r_{L}\sqrt{D}+(m-u)$ both terms are non-negative
for $u_{\mathrm{CJ}}\le u\le m$, and they vanish together only if
$u_{\mathrm{CJ}}=m$, the degenerate normal CJ wave at
$M=M_{\mathrm{CJ}}$.
Table~\ref{tab:polar-classification} collects the resulting
classification.

\begin{table}
 \centering
 \begin{tabular}{lcc}
  Segment & Range & $M_{2}$\\[3pt]
  weak high-compression, CJ point to sonic point &
  $\beta_{\mathrm{CJ}}\le\beta<\beta_{s}$,\ \
  $\theta_{\mathrm{CJ}}\le\theta<\theta_{s}$ & $>1$\\
  weak high-compression, sonic point to detachment &
  $\beta_{s}<\beta\le\beta_{d}$,\ \
  $\theta_{s}<\theta\le\theta_{\max}$ & $<1$\\
  strong high-compression &
  $\beta_{d}<\beta\le\pi/2$ & $<1$\\
  low-compression &
  $\beta_{\mathrm{CJ}}\le\beta\le\pi/2$ & $>1$\\
 \end{tabular}
 \caption{Classification of the equilibrium polar by downstream Mach
 number, for fixed $(\gamma,Q)$ and $M>M_{\mathrm{CJ}}$. $M_{2}=1$
 occurs only at the sonic state $(\beta_{s},\theta_{s})$.}
 \label{tab:polar-classification}
\end{table}
\FloatBarrier

In steady flow, disturbances can propagate upstream through subsonic regions.
The segments with $M_{2}<1$ are therefore susceptible to downstream influence.
This includes the attached weak solution between $\theta_{s}$ and
$\theta_{\max}$, as well as the entire strong branch.

The subsonic part of the weak branch is very narrow. For $\gamma=1.3$
and $Q=10$, the interval between $\theta_{s}$ and $\theta_{\max}$ is
widest near $M=4.19$, where it spans only $0.0126^{\circ}$. The sonic
and detachment criteria therefore give almost the same limiting
deflection.

\section{Extension to \texorpdfstring{two-$\gamma$}{two-gamma} model}
\label{sec:two-gamma}

\subsection{\texorpdfstring{Two-$\gamma$}{Two-gamma} model and governing equations}
\label{sec:two-gamma-model}

The single-$\gamma$ model in \S\ref{sec:model} assigns the same constant
specific-heat ratio to the upstream reactants and the fully reacted products.
The two-$\gamma$ model instead assigns them separate constants,
$\gamma_1>1$ and $\gamma_2>1$, while retaining the same steady planar front
and instantaneous, complete heat release.

Let $u_{ni}$ and $u_{ti}$ denote the normal and tangential velocity
components on side $i$, respectively. Conservation of mass and normal
momentum across the front gives
\begin{equation}
 \rho_1u_{n1}=\rho_2u_{n2},
 \qquad
 p_1+\rho_1u_{n1}^2=p_2+\rho_2u_{n2}^2.
 \label{eq:two-gamma-mass-momentum}
\end{equation}
Conservation of mass and tangential momentum gives
\begin{equation}
 u_{t1}=u_{t2},
 \qquad
 U_1\cos\beta=U_2\cos(\beta-\theta),
 \label{eq:two-gamma-tangential}
\end{equation}
where $U_1$ and $U_2$ are the upstream and downstream speeds.
The tangential kinetic energy thus cancels from the energy balance,
which takes the form
\begin{equation}
 \frac{\gamma_1}{\gamma_1-1}\frac{p_1}{\rho_1}
 +\frac{u_{n1}^2}{2}+Q^*
 =\frac{\gamma_2}{\gamma_2-1}\frac{p_2}{\rho_2}
 +\frac{u_{n2}^2}{2}.
 \label{eq:two-gamma-energy}
\end{equation}
Here $Q^*>0$ denotes the prescribed chemical energy released
per unit mass of the reacting mixture.

Using $u_{n1}=U_1\sin\beta$ and
$u_{n2}=U_2\sin(\beta-\theta)$, mass conservation and
\eqref{eq:two-gamma-tangential} give the geometric relation
\begin{equation}
 r=\frac{\rho_1}{\rho_2}
 =\frac{u_{n2}}{u_{n1}}
 =\frac{\tan(\beta-\theta)}{\tan\beta}.
 \label{eq:two-gamma-geometry}
\end{equation}
As in \S\ref{sec:model}, the dimensionless heat release and
upstream Mach number are defined by
\begin{equation}
 Q=\frac{Q^*}{p_1/\rho_1}=\frac{Q^*}{R_1T_1},
 \qquad
 M=\frac{U_1}{\sqrt{\gamma_1p_1/\rho_1}},
 \label{eq:two-gamma-heat}
\end{equation}
where $R_1$ is the specific gas constant of the upstream mixture.

The dimensional mass and momentum conditions
\eqref{eq:two-gamma-mass-momentum}--\eqref{eq:two-gamma-tangential}
contain no explicit specific-heat ratio, and therefore retain their
single-$\gamma$ forms. The geometric relation
\eqref{eq:two-gamma-geometry} likewise coincides with
\eqref{eq:reciprocal-density-ratio}. The change appears in the thermal
enthalpy terms of \eqref{eq:two-gamma-energy}: the common coefficient
$\gamma/(\gamma-1)$ is replaced by $\gamma_1/(\gamma_1-1)$ upstream
and $\gamma_2/(\gamma_2-1)$ downstream. These coefficients become equal
when $\gamma_1=\gamma_2$, recovering the single-$\gamma$ model.

\subsection{Reduction to the \texorpdfstring{single-$\gamma$}{single-gamma} model}
\label{sec:two-gamma-reduction}

To recover a common enthalpy coefficient in
\eqref{eq:two-gamma-energy}, we combine the upstream thermal enthalpy
and the prescribed heat release into an equivalent expression using
the downstream coefficient:
\begin{equation}
 \frac{\gamma_1}{\gamma_1-1}\frac{p_1}{\rho_1}+Q^*
 =\frac{\gamma_2}{\gamma_2-1}\frac{p_1}{\rho_1}+Q_{\mathrm e}^*,
 \label{eq:two-gamma-enthalpy-rewrite}
\end{equation}
where
\begin{equation}
 Q_{\mathrm e}^*=Q^*
 +\left(\frac{\gamma_1}{\gamma_1-1}
       -\frac{\gamma_2}{\gamma_2-1}\right)\frac{p_1}{\rho_1}.
 \label{eq:two-gamma-effective-heat-dimensional}
\end{equation}
The energy equation therefore becomes
\begin{equation}
 \frac{\gamma_2}{\gamma_2-1}\frac{p_1}{\rho_1}
 +\frac{u_{n1}^2}{2}+Q_{\mathrm e}^*
 =\frac{\gamma_2}{\gamma_2-1}\frac{p_2}{\rho_2}
 +\frac{u_{n2}^2}{2}.
 \label{eq:two-gamma-effective-energy}
\end{equation}
Together with \eqref{eq:two-gamma-mass-momentum} and
\eqref{eq:two-gamma-tangential}, this is precisely the single-$\gamma$
jump problem with specific-heat ratio $\gamma_2$ and effective heat release
$Q_{\mathrm e}^*$. The geometric relation
\eqref{eq:two-gamma-geometry} also carries over directly.

To use the preceding dimensionless formulae, the upstream speed must also
be expressed using the sound-speed scale of this equivalent model.
The required effective parameters are
\begin{equation}
 \boxed{\begin{aligned}
 \gamma_{\mathrm e}&=\gamma_2,\\
 M_{\mathrm e}&=\frac{U_1}{\sqrt{\gamma_2p_1/\rho_1}}
              =M\sqrt{\frac{\gamma_1}{\gamma_2}},\\
 Q_{\mathrm e}&=\frac{Q_{\mathrm e}^*}{p_1/\rho_1}
              =Q+\frac{\gamma_1}{\gamma_1-1}
                 -\frac{\gamma_2}{\gamma_2-1}.
 \end{aligned}}
 \label{eq:two-gamma-map}
\end{equation}
Here $M$ and $Q$ remain the actual upstream Mach number and the prescribed
dimensionless chemical energy release. The effective quantities are
parameters of the equivalent single-$\gamma$ problem. In particular,
$Q_{\mathrm e}$ includes the difference between the two thermal enthalpy
coefficients at the upstream state.

The reduction is exact.
It preserves the physical pressures, densities and velocities, and hence
also $r$, $\theta$ and $\beta$. Consequently, the wave-angle cubic
\eqref{eq:cubic}--\eqref{eq:cubic-coefficients} applies directly after
the substitution
\begin{equation}
 (\gamma,M,Q)\ \longmapsto\
 (\gamma_2,M_{\mathrm e},Q_{\mathrm e}).
 \label{eq:two-gamma-substitution}
\end{equation}
Its candidate roots can be evaluated using the same closed-form expressions
as in \S\ref{sec:selection}. 
When $\gamma_1=\gamma_2=\gamma$, \eqref{eq:two-gamma-map} gives
$M_{\mathrm e}=M$ and $Q_{\mathrm e}=Q$, recovering the original model.

\subsection{Physical branches and critical points}
\label{sec:two-gamma-branches}

For $Q_{\mathrm e}>0$ and
$M_{\mathrm e}>M_{\mathrm{CJ}}(\gamma_2,Q_{\mathrm e})$, the
detonation-branch classification in \S\ref{sec:selection} applies at the
effective parameters. Since the mapping preserves $\beta$, $\theta$ and
$r$, the geometric admissibility conditions and the high- and
low-compression classification retain their physical meaning in the
two-$\gamma$ model. The sign criterion \eqref{eq:branch-sign}, evaluated
with the substitution \eqref{eq:two-gamma-substitution}, identifies the
compression branch of each candidate root. The weak and strong
overdriven solutions are then selected by the same wave-angle ordering
as in the single-$\gamma$ model.

The CJ endpoint, detachment point and downstream total-sonic point are
also obtained by the same substitution in
\eqref{eq:u-cj}--\eqref{eq:theta-cj},
\eqref{eq:detachment-cubic}--\eqref{eq:detachment-angles} and
\eqref{eq:sonic-root}--\eqref{eq:sonic-deflection}, respectively.
In particular, the downstream Mach number
$M_2=U_2/\sqrt{\gamma_2p_2/\rho_2}$ is unchanged because both the
downstream state and its specific-heat ratio are preserved.
The sonic point therefore retains its role in separating the
downstream-supersonic and downstream-subsonic portions of the weak branch,
and the strong branch remains subsonic downstream.
All resulting wave and deflection angles apply directly to the original
two-$\gamma$ model; a critical upstream Mach number expressed in the
effective variables is converted back using
$M=M_{\mathrm e}\sqrt{\gamma_2/\gamma_1}$.

\section{Application to chemical-equilibrium calculations}
\label{sec:chemical-equilibrium}

We now apply the closed-form solution to equilibrium oblique detonations
in chemically reacting mixtures. In this setting, the product composition
is determined by chemical equilibrium, and the species specific heats
vary with temperature. The wave angle and downstream state are obtained
by coupling the conservation relations across the detonation wave to the
calculation of the equilibrium composition.

\subsection{Closed-form integration into an iterative chemical-equilibrium framework}
\label{sec:equilibrium-framework}

Chemical reactions in oblique detonations can be treated by resolving
the finite-rate reaction process or by directly calculating the
chemical-equilibrium state. The former approach couples the flow
equations with chemical kinetics to describe the evolution of the
reacting mixture through the reaction zone. The latter determines the
downstream equilibrium state by iteratively solving the conservation
relations together with the chemical-equilibrium conditions, without
resolving the reaction-zone structure. Here we adopt the latter approach
to calculate the equilibrium wave angle and downstream state for a
prescribed upstream state and deflection angle.

\citet{zhang2022equilibrium} developed a two-step iterative scheme for
solving shock relations coupled with chemical equilibrium, with
applications to oblique detonations. In the first step, the upstream
state and downstream composition are held fixed, and the conservation
and geometric relations are reduced to a single nonlinear equation
for the wave angle. Newton iteration gives the wave angle, from which
the downstream temperature and pressure are obtained. In the second
step, the equilibrium composition at this temperature and pressure is
calculated by minimising the Gibbs free energy subject to elemental
conservation, following the NASA formulation of
\citet{gordon1994cea}. The downstream composition is then updated with
relaxation, and the two steps are repeated until convergence.

We retain this two-step scheme to calculate equilibrium oblique
detonations and replace the Newton iteration for the wave angle in
the first step with the closed-form formulae. The connection to the preceding analysis is made
by locally linearising the mixture enthalpy at each iteration.
This gives the enthalpy form used in \S\ref{sec:two-gamma-reduction},
so the same effective-parameter reduction and explicit cubic solution
can be applied.

Let $T_2^{(k)}$ and $\boldsymbol X_2^{(k)}$ denote the downstream
temperature and species mole-fraction vector at iteration $k$.
At this composition, the mixture specific enthalpy is approximated by
\begin{equation}
 h_2(T,\boldsymbol X_2^{(k)})
 \simeq h_2^{(k)}+c_{p,2}^{(k)}(T-T_2^{(k)}),
 \label{eq:equilibrium-linear-enthalpy}
\end{equation}
where $h_2^{(k)}=h_2(T_2^{(k)},\boldsymbol X_2^{(k)})$.
The coefficient
$c_{p,2}^{(k)}=(\partial h_2/\partial T)_{\boldsymbol X_2}$,
evaluated at the current state, is the specific heat at constant
pressure and fixed composition. The mixture enthalpy includes the
species enthalpies of formation, accounting for the chemical energy
difference between reactants and products. With
$R_2^{(k)}=R(\boldsymbol X_2^{(k)})$, the ideal-gas relation
$T=p/(\rho R_2^{(k)})$ makes the linearised enthalpy a term proportional
to $p/\rho$ plus a constant. Matching these terms to
\eqref{eq:two-gamma-effective-energy} gives
\begin{equation}
 \begin{aligned}
 \gamma_{\mathrm e}^{(k)}
 &=\frac{c_{p,2}^{(k)}}{c_{p,2}^{(k)}-R_2^{(k)}},\\
 M_{\mathrm e}^{(k)}
 &=\frac{U_1}{\sqrt{\gamma_{\mathrm e}^{(k)}p_1/\rho_1}},\\
 Q_{\mathrm e}^{(k)}
 &=\frac{h_1-h_2^{(k)}+c_{p,2}^{(k)}T_2^{(k)}}{p_1/\rho_1}
   -\frac{c_{p,2}^{(k)}}{R_2^{(k)}}.
 \end{aligned}
 \label{eq:equilibrium-effective-parameters}
\end{equation}
Here $h_1=h_1(T_1,\boldsymbol X_1)$ is fixed by the upstream state.
Using these effective parameters, the wave-angle cubic
\eqref{eq:cubic} is solved explicitly at the prescribed deflection
angle $\theta$. Its admissible roots are classified on the polar of
the local linearised model. When both high-compression roots exist,
the smaller and larger wave angles give the weak and strong solutions
of that model, respectively.

For the selected root, the density ratio follows from
\eqref{eq:two-gamma-geometry}. The intermediate downstream pressure
and temperature are then
\begin{equation}
 \widetilde p_2
 =p_1+\rho_1U_1^2\sin^2\beta\,(1-r),
 \qquad
 \widetilde T_2
 =\frac{\widetilde p_2\,r}{\rho_1R_2^{(k)}}.
 \label{eq:equilibrium-flow-update}
\end{equation}
At $(\widetilde T_2,\widetilde p_2)$, Gibbs free-energy minimisation
subject to the elemental composition of the reactants gives the
equilibrium mole-fraction vector $\boldsymbol X_{2,\mathrm{eq}}$.
The composition and temperature are updated according to
\begin{equation}
 \boldsymbol X_2^{(k+1)}
 =(1-\lambda)\boldsymbol X_2^{(k)}
 +\lambda\boldsymbol X_{2,\mathrm{eq}},
 \qquad T_2^{(k+1)}=\widetilde T_2,
 \label{eq:equilibrium-outer-update}
\end{equation}
where $0<\lambda\leq1$ is the composition relaxation factor.
The updated state defines the enthalpy linearisation for the next
iteration.

These steps are repeated until the flow state and composition converge.
At convergence, the linearised enthalpy coincides with the full mixture
enthalpy at the resulting temperature and composition. The closed-form
formulae thus replace the inner wave-angle iteration while retaining
the coupling between the conservation relations and chemical
equilibrium.

\subsection{Numerical verification and computational efficiency}
\label{sec:equilibrium-verification}

We compare two solvers for oblique detonations with chemically
equilibrated products.
The first follows the two-step scheme of \citet{zhang2022equilibrium} and determines
the wave angle by Newton iteration. The second retains this scheme but
replaces the wave-angle calculation with the closed-form formulae
introduced in \S\ref{sec:equilibrium-framework}. Both solvers are
implemented in Python and use the same thermodynamic data,
chemical-equilibrium solver and convergence tolerances.

The calculations consider stoichiometric hydrogen--air, methane--air
and ethylene--air mixtures at $T_1=300\,\mathrm{K}$ and
$p_1=1\,\mathrm{atm}$, with upstream Mach numbers $M=6,7,8,9,10$.
For each condition, ten deflection angles are selected according to
$\theta=\theta_{\mathrm{CJ}}
+f(\theta_{\max}-\theta_{\mathrm{CJ}})$, where
$f=0.05,0.15,\ldots,0.95$. The main test set therefore contains
150 cases, each evaluated on both the weak and strong branches.
Additional sampling is performed near detachment.

\begin{figure}[!ht]
 \centering
 \includegraphics[width=\linewidth]{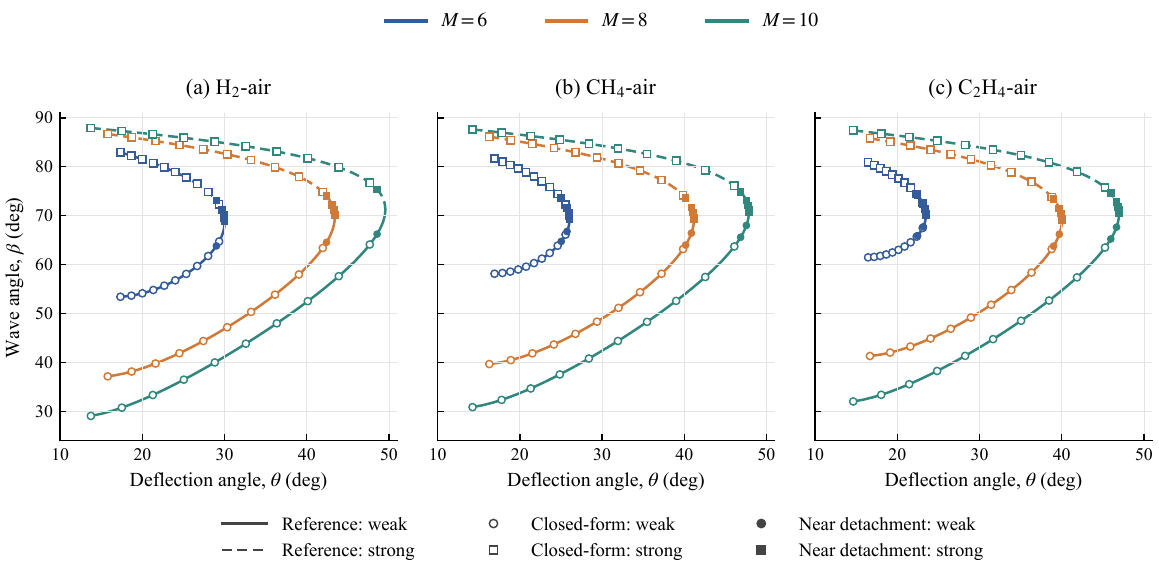}
 \caption{Wave angle versus deflection angle for stoichiometric
 (a) hydrogen--air, (b) methane--air and (c) ethylene--air mixtures
 at $T_1=300\,\mathrm{K}$ and $p_1=1\,\mathrm{atm}$.
 Colours indicate $M=6,8,10$.
 Solid and dashed curves show the weak and strong reference solutions.
 Circles and squares show the corresponding solutions obtained with
 the closed-form formulation; filled symbols denote the additional
 converged solutions from refined sampling near detachment.}
 \label{fig:equilibrium-three-gas}
\end{figure}
\FloatBarrier

Figure~\ref{fig:equilibrium-three-gas} shows the results at $M=6,8,10$.
The reference polars are obtained by numerically solving the
Rankine--Hugoniot relations under chemical equilibrium, with the
downstream pressure as the prescribed parameter. The symbols represent
solutions obtained with the closed-form formulation at prescribed
deflection angles, with additional sampling near detachment.
The calculated wave angles agree with the reference curves on both
the weak and strong branches for all three mixtures.

The two solvers also give consistent results for the 300 branch states
in the main test set. Their maximum relative difference in wave angle,
downstream temperature and pressure is below $1.2\times10^{-10}$,
and the maximum absolute difference in species mole fractions is below
$4.0\times10^{-11}$.

Table~\ref{tab:equilibrium-timing} compares the elapsed times
and the numbers of chemical-equilibrium evaluations for the
main test set. The timings include both the flow-state and
chemical-equilibrium calculations and are reported as medians
over seven repeated runs.

\begin{table}
\centering
\begin{tabular}{lrrrrr}
 & \multicolumn{2}{c}{Elapsed time (s)}
 & \multicolumn{2}{c}{Equilibrium evaluations}
 & \\ 
Mixture & Newton-based & Closed-form
        & Newton-based & Closed-form & Speedup \\[3pt]
Hydrogen--air & 1.841 & 1.387 & 4311 & 4309 & 1.32 \\
Methane--air  & 2.489 & 1.734 & 4304 & 4304 & 1.44 \\
Ethylene--air & 2.543 & 1.760 & 4311 & 4312 & 1.45 \\
All cases    & 6.872 & 4.875 & 12926 & 12925 & 1.41 \\
\end{tabular}
\caption{Computational cost for the main test set.
Evaluation counts are totals for one run, including both
branches. Speedups are the median ratios of Newton-based
to closed-form solver times over the seven runs.}
\label{tab:equilibrium-timing}
\end{table}

The closed-form solver is faster for all three mixtures,
reducing the overall elapsed time by approximately $29\%$.
The two solvers require nearly identical numbers of
chemical-equilibrium evaluations, indicating that the saving
comes primarily from replacing the inner wave-angle iteration
with a closed-form evaluation.

\section{Open-source implementation}
\label{sec:implementation}

A Python implementation accompanies this work, covering the
single-$\gamma$ model, the two-$\gamma$ extension and the
chemical-equilibrium application. The closed-form core provides
wave-angle solutions with physical branch selection, together
with the CJ, detachment and downstream sonic points.
The two-$\gamma$ model uses the same core through the exact
parameter mapping introduced in \S\ref{sec:two-gamma-reduction}.
For chemically equilibrated products, the core is incorporated
into the two-step iterative scheme described in
\S\ref{sec:equilibrium-framework}, replacing the inner wave-angle
iteration while retaining the coupling between the flow state
and product composition.

The source code is available at
\url{https://github.com/lijing-creator/Exact_explicit_wave_angle_solutions}.
The repository includes documentation, worked examples,
implementation checks against independent numerical solutions,
and scripts and data for reproducing the principal numerical
results in this paper. Installation instructions,
model conventions and interface details are provided in the
repository documentation.

\section{Conclusions}
\label{sec:conclusions}
For a prescribed deflection, the equilibrium wave angles are the roots of a
cubic in $\tan\beta$. Heat release enters only through an additive term, so the
cubic reduces to the classical oblique-shock cubic at $Q=0$. It is therefore
solved by the same route, Cardano's method. At most two of the three roots
satisfy the attachment condition. The elimination that produces the cubic
removes the density ratio, and with it the branch identity of each root. The
branch is recovered from the sign of $\sigma=1+\gamma u-(\gamma+1)ur$: $\sigma$
is positive on the high-compression root, where it equals $+\sqrt{D}$, and
negative on the low-compression root, where it equals $-\sqrt{D}$. The weak
overdriven solution therefore exists precisely on
$\theta_{\mathrm{CJ}}<\theta<\theta_{\max}$, while below
$\theta_{\mathrm{CJ}}$ the two attached roots are the strong and the
low-compression solutions. At detachment the two attached roots merge, and the
double root satisfies a cubic of the same type. Cardano's method applies to it
as well, and at $Q=0$ it reduces to the standard oblique-shock detachment
quadratic.

The downstream total-sonic condition reduces to a quadratic equation in
$M^{2}\sin^{2}\beta$, with exactly one physical root on each attached
high-compression polar. The CJ, sonic and detachment deflections satisfy the
strict ordering
\begin{equation*}
 \theta_{\mathrm{CJ}}<\theta_{s}<\theta_{\max},
\end{equation*}
where $\theta_{\mathrm{CJ}}$, $\theta_{s}$ and $\theta_{\max}$ denote
the deflection angles at the CJ endpoint, downstream total-sonic point and
detachment point, respectively. Along the weak branch, the
downstream flow is supersonic between the CJ endpoint and the
total-sonic point, and subsonic between the total-sonic point and
detachment. The strong branch is subsonic throughout, and the two
branches meet at a subsonic detachment state. Thus the total-sonic point divides
the weak branch into downstream-supersonic and downstream-subsonic parts.

The closed-form analysis assumes a calorically perfect gas
with fixed heat release $Q$. Heat release is instantaneous and complete,
and the oblique detonation is represented by a planar, zero-thickness
equilibrium discontinuity attached to a straight wedge.
The closed-form relations determine the equilibrium states and
the structure of the polar, but do not represent finite-thickness induction
and reaction zones, curved transition structures
\citep{li1994,papalexandris2000}, or multidimensional and cellular
instabilities \citep{teng2014}.
The two-$\gamma$ model can be reduced to the single-$\gamma$
model through an exact parameter mapping.
Within the same equilibrium model,
these relations can be used directly in method-of-characteristics
calculations and in preliminary screening of attached wedge
configurations. For prescribed upstream conditions and deflection,
the closed-form relations give the admissible wave angles and downstream
Mach regime directly. No iterative inversion of the polar is required.
These relations can also be used in oblique-detonation calculations
with equilibrium chemistry, where they replace the inner wave-angle
iteration and reduce the computational cost.

\begin{bmhead}[Funding.]
This work was supported by the National Natural Science Foundation of China
(grant numbers 12072353, 12132017).
\end{bmhead}

\begin{bmhead}[Declaration of interests.]
The authors report no conflict of interest.
\end{bmhead}

\begin{bmhead}[Data availability statement.]
The code implementing the closed-form relations, together with the
verification and figure-generation scripts, is openly available at
\url{https://github.com/lijing-creator/Exact_explicit_wave_angle_solutions}.
\end{bmhead}

\begin{bmhead}[Author ORCIDs.]
J. Li, https://orcid.org/0009-0001-7568-0341;
C. Luo, https://orcid.org/0000-0002-6283-1817.
\end{bmhead}

\begin{bmhead}[Author contributions.]
J.L. conceived the study, developed the analysis, wrote the code, carried out
the verification and prepared the original draft. C.L. supervised the work,
acquired the funding, and reviewed and edited the manuscript.
\end{bmhead}

\appendix
\renewcommand{\theHsection}{appendix.\Alph{section}}
\section{Explicit radical roots of the wave-angle cubic}
\label{app:cubic-roots}

The three roots of \eqref{eq:cubic} are recorded here without derivation,
directly in terms of $M$, $\theta$, $\gamma$ and $Q$. Define
\begin{equation}
 \begin{aligned}
  \Delta_{0}={}&\tan^{2}\theta\Biggl\{
  4\left[\frac{2(\gamma-1)}{\gamma}Q-M^{2}+1\right]^{2}\\
  &-3\left[\frac{2(\gamma-1)}{\gamma}Q
  +(\gamma-1)M^{2}+2\right]\\
  &\hspace{12mm}\times
  \left\{\frac{2(\gamma-1)}{\gamma}Q
  +\tan^{2}\theta\left[(\gamma+1)M^{2}+2\right]\right\}\Biggr\},
 \end{aligned}
 \label{eq:app-delta-zero}
\end{equation}
and
\begin{equation}
 \begin{aligned}
  \Delta_{1}={}&\tan^{3}\theta\Biggl\{
  16\left[\frac{2(\gamma-1)}{\gamma}Q-M^{2}+1\right]^{3}\\
  &-18\left[\frac{2(\gamma-1)}{\gamma}Q
  +(\gamma-1)M^{2}+2\right]
  \left[\frac{2(\gamma-1)}{\gamma}Q-M^{2}+1\right]\\
  &\hspace{12mm}\times
  \left\{\frac{2(\gamma-1)}{\gamma}Q
  +\tan^{2}\theta\left[(\gamma+1)M^{2}+2\right]\right\}\\
  &+54\tan^{2}\theta
  \left[\frac{2(\gamma-1)}{\gamma}Q
  +(\gamma-1)M^{2}+2\right]^{2}\Biggr\}.
 \end{aligned}
 \label{eq:app-delta-one}
\end{equation}
The associated radical quantity and cube root of unity are
\begin{equation}
 \begin{gathered}
  \Lambda=\Delta_{1}^{2}-4\Delta_{0}^{3}
  =2916\tan^{12}\theta
  \left[\frac{2(\gamma-1)}{\gamma}Q
  +(\gamma-1)M^{2}+2\right]^{6}\Delta_{C},\\
  \omega=\exp(2\pi\mathrm{i}/3).
 \end{gathered}
 \label{eq:app-radical-quantities}
\end{equation}
For $(\Delta_{0},\Delta_{1})\ne(0,0)$, choose the sign for which
\begin{equation}
 \mathcal{C}
 =\sqrt[3]{\frac{\Delta_{1}\pm\sqrt{\Lambda}}{2}}
 \ne0.
 \label{eq:app-cardano-C}
\end{equation}
Any one of the three values of this cube root may be used. The three
algebraic roots are then
\begin{equation}
 \begin{aligned}
  t_{k}={}&-\frac{1}{
  3\tan^{2}\theta
  \left[\dfrac{2(\gamma-1)}{\gamma}Q+(\gamma-1)M^{2}+2\right]}\\
  &\times\left\{
  2\tan\theta\left[\frac{2(\gamma-1)}{\gamma}Q-M^{2}+1\right]
  +\omega^{k}\mathcal{C}
  +\frac{\Delta_{0}}{\omega^{k}\mathcal{C}}\right\},\\
  k=0,1,2.
 \end{aligned}
 \label{eq:app-cardano-roots}
\end{equation}

When $\Lambda=0$ and $\Delta_{0}\ne0$, the simple and double roots are
\begin{equation}
 \begin{aligned}
  t_{s}&=-\frac{
  2\tan\theta\left[\dfrac{2(\gamma-1)}{\gamma}Q-M^{2}+1\right]
  \Delta_{0}+\Delta_{1}}
  {3\tan^{2}\theta
  \left[\dfrac{2(\gamma-1)}{\gamma}Q+(\gamma-1)M^{2}+2\right]
  \Delta_{0}},\\[2mm]
  t_{d}&=\frac{
  \Delta_{1}-4\tan\theta
  \left[\dfrac{2(\gamma-1)}{\gamma}Q-M^{2}+1\right]\Delta_{0}}
  {6\tan^{2}\theta
  \left[\dfrac{2(\gamma-1)}{\gamma}Q+(\gamma-1)M^{2}+2\right]
  \Delta_{0}}.
 \end{aligned}
 \label{eq:app-repeated-roots}
\end{equation}
When $\Delta_{0}=\Delta_{1}=0$, all three roots reduce to
\begin{equation}
 t_{*}=-\frac{
 2\left[\dfrac{2(\gamma-1)}{\gamma}Q-M^{2}+1\right]}
 {3\tan\theta
 \left[\dfrac{2(\gamma-1)}{\gamma}Q+(\gamma-1)M^{2}+2\right]}.
 \label{eq:app-triple-root}
\end{equation}
Only the real roots satisfying the filter in \S\ref{sec:selection} are
retained as physical wave-angle candidates.

\section{Bounds and branch geometry of the attached polar}
\label{app:polar-geometry}

This appendix proves the branch geometry stated at the end of
\S\ref{sec:detachment}. Fix $\gamma>1$, $Q>0$ and
$M>M_{\mathrm{CJ}}$, let $(y_{d},s_{d})$ denote the admissible detachment
pair, and parameterise the two equilibrium branches by
$\beta\in[\beta_{\mathrm{CJ}},\pi/2]$:
\begin{equation}
 u=M^{2}\sin^{2}\beta,\qquad
 \theta_{H,L}(\beta)=\beta-\arctan\!\bigl[r_{H,L}(u)\tan\beta\bigr].
 \label{eq:app-branch-parameterisation}
\end{equation}
Since $0\le\sqrt{D}<u-1$ for $u\ge u_{\mathrm{CJ}}>1$,
\eqref{eq:density-roots} gives the density-ratio bounds
\begin{equation}
 0<r_{H}\le r_{L}<1,
 \label{eq:app-density-bounds}
\end{equation}
hence $0<\theta_{H,L}(\beta)<\beta$ for $\beta<\pi/2$. Both functions are
continuous, with
$\theta_{H,L}(\beta_{\mathrm{CJ}})=\theta_{\mathrm{CJ}}$,
$\theta_{H}(\beta_{d})=\theta_{\max}$ and $\theta_{H,L}(\pi/2)=0$.
The proofs below use three facts: each point on either branch solves the
cubic \eqref{eq:cubic} at its own deflection; the product relation in
\eqref{eq:cubic-vieta} leaves at most two positive roots; the two branches
intersect only at $\beta_{\mathrm{CJ}}$, where $r_{H}=r_{L}$.

The double root is interior, $u_{d}>u_{\mathrm{CJ}}$. To see this, factor
the cubic as
\begin{equation*}
 C_{3}(t)=t(1+\tau t)^{2}\,\Phi\bigl(r(t),u(t)\bigr),
\end{equation*}
with $\Phi(r,u)$ the left-hand side of \eqref{eq:rh-quadratic} and $r(t)$,
$u(t)$ from \eqref{eq:tangent-geometry}. A root with $u=u_{\mathrm{CJ}}$
has $r=r_{\mathrm{CJ}}$ and $\Phi_{r}=0$. The other partial derivative is
\begin{equation*}
 \Phi_{u}=(\gamma+1)(r-1)\bigl[r-(\gamma-1)/(\gamma+1)\bigr],
\end{equation*}
which is negative at $r_{\mathrm{CJ}}\in\bigl(\gamma/(\gamma+1),1\bigr)$,
and $u'(t)>0$. Hence $C_{3}'=t(1+\tau t)^{2}\Phi_{u}u'\ne0$, so every
root with $u=u_{\mathrm{CJ}}$ is simple. The detachment double root
therefore satisfies $u_{d}>u_{\mathrm{CJ}}$, and the high-compression
branch splits at $\beta_{d}>\beta_{\mathrm{CJ}}$ into a weak segment
$\beta\in[\beta_{\mathrm{CJ}},\beta_{d}]$ and a strong segment
$\beta\in[\beta_{d},\pi/2]$.

$\theta_{\max}$ is the strict maximum of $\theta_{H}$, attained only at
$\beta_{d}$. The only positive root at $\theta=\theta_{\max}$ is $t_{d}$
(\S\ref{sec:selection}), so the branch meets this level only at
$\beta_{d}$. Writing \eqref{eq:cubic-tau-t} as $C_{3}(t;\tau)$, the
recovery relation \eqref{eq:s-recovery} gives
\begin{equation}
 \left.\frac{\partial C_{3}}{\partial\tau}\right|_{(t_{d},\tau_{d})}
 =\frac{2}{s_{d}}
 \bigl(Ay_{d}^{3}+By_{d}^{2}+Cs_{d}y_{d}+s_{d}\bigr)
 =2\,(Cy_{d}+2)>0.
 \label{eq:polar-tangency-sign}
\end{equation}
At $\tau=\tau_{d}$ the cubic factors as
\begin{equation*}
 C_{3}(t;\tau_{d})=As_{d}(t-t_{d})^{2}(t-t_{s}),
 \qquad t_{s}<0,
\end{equation*}
where the sign of $t_{s}$ follows from the product relation in
\eqref{eq:cubic-vieta}. Hence $C_{3}(t;\tau_{d})>0$ for $t\ne t_{d}$
sufficiently close to $t_{d}$. A nearby point on the branch satisfies
$C_{3}(t;\tau)=0$, so \eqref{eq:polar-tangency-sign} and the mean-value
theorem in $\tau$ give $\tau<\tau_{d}$.

The difference $\theta_{H}-\theta_{\max}$ therefore never vanishes for
$\beta\ne\beta_{d}$, and it is negative near $\beta_{d}$. By continuity,
$\theta_{H}(\beta)<\theta_{\max}$ for all $\beta\ne\beta_{d}$, and in
particular $\theta_{\mathrm{CJ}}<\theta_{\max}$.

Along the weak segment $\beta\in[\beta_{\mathrm{CJ}},\beta_{d}]$, the
deflection increases strictly. Fix $\hat{\theta}$ with
$0<\hat{\theta}<\theta_{\max}$. The strong segment
$\beta\in[\beta_{d},\pi/2]$ runs continuously from $\theta_{\max}$ to $0$,
so by the intermediate value theorem it already supplies one positive root
at this deflection, at a wave angle exceeding $\beta_{d}$. The product
relation in \eqref{eq:cubic-vieta} allows at most two positive roots, so
the weak segment supplies at most one. Hence $\theta_{H}$ is continuous
and injective on the weak segment, and therefore strictly monotone. Its endpoint values
increase from $\theta_{\mathrm{CJ}}$ to $\theta_{\max}$, so it is
strictly increasing, and the intermediate value theorem places the weak
overdriven solution precisely on
$\theta_{\mathrm{CJ}}<\theta<\theta_{\max}$.

Beyond the CJ point the low-compression branch stays below
$\theta_{\mathrm{CJ}}$. At $\theta=\theta_{\mathrm{CJ}}$ the two branches
meet at $\beta_{\mathrm{CJ}}$, which gives one positive root, and since
$\theta_{\mathrm{CJ}}<\theta_{\max}$ the strong segment supplies a second
at a wave angle exceeding $\beta_{d}$. The product relation in
\eqref{eq:cubic-vieta} admits no third, and the two branches meet only at
$\beta_{\mathrm{CJ}}$, so
$\theta_{L}(\beta)\ne\theta_{\mathrm{CJ}}$ for $\beta>\beta_{\mathrm{CJ}}$.
Thus $\theta_{L}-\theta_{\mathrm{CJ}}$ never vanishes on
$(\beta_{\mathrm{CJ}},\pi/2]$, and $\theta_{L}(\pi/2)=0$ fixes its sign.
For $0<\theta<\theta_{\mathrm{CJ}}$, both the low-compression branch and
the strong segment attain the prescribed deflection by the intermediate
value theorem. By the product relation in \eqref{eq:cubic-vieta} at most
two roots are positive, so these are the only two.

\section{Signs of the sonic-quadratic coefficients}
\label{app:sonic-coefficient-signs}

Let $\gamma>1$, $Q>0$ and $M>M_{\mathrm{CJ}}$. The CJ relation
\begin{equation*}
 (u_{\mathrm{CJ}}-1)^{2}
 =(\gamma+1)q'u_{\mathrm{CJ}}
\end{equation*}
implies
\begin{equation*}
 \begin{aligned}
 u_{\mathrm{CJ}}-1-(\gamma+1)q'
 &=(u_{\mathrm{CJ}}-1)
 -\frac{(u_{\mathrm{CJ}}-1)^{2}}{u_{\mathrm{CJ}}}\\
 &=\frac{u_{\mathrm{CJ}}-1}{u_{\mathrm{CJ}}}>0.
 \end{aligned}
\end{equation*}
Since $m>u_{\mathrm{CJ}}$, it follows that
\begin{equation*}
 m-1>u_{\mathrm{CJ}}-1>(\gamma+1)q'.
\end{equation*}
Using \eqref{eq:E-coefficients}, we therefore obtain
\begin{equation*}
 \begin{aligned}
 c_{2}
 &<\gamma\bigl[(\gamma+2)q'-2(\gamma+1)q'\bigr]
 =-\gamma^{2}q'<0,\\
 c_{0}
 &>2(\gamma+1)q'-q'=(2\gamma+1)q'>0.
 \end{aligned}
\end{equation*}
For $c_{1}$, no further CJ bound is needed. Indeed, its first term is
non-negative, whereas its second term is strictly positive:
\begin{equation*}
 (\gamma+1)(m-1-q')^{2}\geq 0,
 \qquad
 2\bigl[(\gamma-1)(m-1)+q'\bigr]>0,
\end{equation*}
because $\gamma>1$, $m-1>0$ and $q'>0$. Their sum is therefore
$c_{1}>0$.
Thus $c_{2}<0$, $c_{1}>0$ and $c_{0}>0$ throughout the stated physical
domain.

\bibliographystyle{jfm}
\bibliography{references}

@article{pratt1991,
  author  = {Pratt, D. T. and Humphrey, J. W. and Glenn, D. E.},
  title   = {Morphology of standing oblique detonation waves},
  journal = {Journal of Propulsion and Power},
  year    = {1991},
  volume  = {7},
  number  = {5},
  pages   = {837--845},
  doi     = {10.2514/3.23399}
}

@article{thompson1950,
  author  = {Thompson, M. J.},
  title   = {A note on the calculation of oblique shock-wave characteristics},
  journal = {Journal of the Aeronautical Sciences},
  year    = {1950},
  volume  = {17},
  number  = {11},
  pages   = {744},
  doi     = {10.2514/8.1790}
}

@article{mascitti1969,
  author  = {Mascitti, V. R.},
  title   = {A closed-form solution to oblique shock-wave properties},
  journal = {Journal of Aircraft},
  year    = {1969},
  volume  = {6},
  number  = {1},
  pages   = {66},
  doi     = {10.2514/3.59421}
}

@techreport{wellmann1972,
  author      = {Wellmann, J.},
  title       = {Vereinfachung von {R}echnungen am schiefen {V}erdichtungssto{\ss}},
  institution = {DFVLR, Institut f{\"u}r Aerodynamik},
  type        = {Forschungsbericht},
  number      = {DLR-FB 72-11},
  address     = {Braunschweig},
  year        = {1972},
  note        = {in German}
}

@techreport{hartley1991,
  author      = {Hartley, T. T. and Brandis, R. and Mossayebi, F.},
  title       = {Exact and approximate solutions to the oblique shock equations for real-time applications},
  institution = {National Aeronautics and Space Administration},
  type        = {NASA Contractor Report},
  number      = {187173},
  year        = {1991}
}

@book{emanuel2001,
  author    = {Emanuel, G.},
  title     = {Analytical Fluid Dynamics},
  edition   = {2},
  publisher = {CRC Press},
  address   = {Boca Raton, FL},
  year      = {2001},
  note      = {Appendix C}
}

@article{powersgonthier1992,
  author  = {Powers, J. M. and Gonthier, K. A.},
  title   = {Reaction zone structure for strong, weak overdriven, and weak underdriven oblique detonations},
  journal = {Physics of Fluids A},
  year    = {1992},
  volume  = {4},
  number  = {9},
  pages   = {2082--2089},
  doi     = {10.1063/1.858378}
}

@article{powersstewart1992,
  author  = {Powers, J. M. and Stewart, D. S.},
  title   = {Approximate solutions for oblique detonations in the hypersonic limit},
  journal = {AIAA Journal},
  year    = {1992},
  volume  = {30},
  number  = {3},
  pages   = {726--736},
  doi     = {10.2514/3.10978}
}

@article{jiang2023,
  author  = {Jiang, Z.},
  title   = {Standing oblique detonation for hypersonic propulsion: A review},
  journal = {Progress in Aerospace Sciences},
  year    = {2023},
  volume  = {143},
  pages   = {100955},
  doi     = {10.1016/j.paerosci.2023.100955}
}

@article{teng2020,
  author  = {Teng, H. H. and Jiang, Z. L.},
  title   = {Progress in multi-wave structure and stability of oblique detonations},
  journal = {Advances in Mechanics},
  year    = {2020},
  volume  = {50},
  pages   = {202002},
  note    = {in Chinese},
  doi     = {10.6052/1000-0992-19-011}
}

@techreport{townend1970,
  author      = {Townend, L. H.},
  title       = {An analysis of oblique and normal detonation waves},
  institution = {Aeronautical Research Council},
  type        = {Reports and Memoranda},
  number      = {3638},
  address     = {London},
  publisher   = {Her Majesty's Stationery Office},
  year        = {1970}
}

@article{gross1963,
  author  = {Gross, R. A.},
  title   = {Oblique detonation waves},
  journal = {AIAA Journal},
  year    = {1963},
  volume  = {1},
  number  = {5},
  pages   = {1225--1227},
  doi     = {10.2514/3.1777}
}

@article{siestrunck1953,
  author  = {Siestrunck, R. and Fabri, J. and Le Griv{\`e}s, E.},
  title   = {Some properties of stationary detonation waves},
  journal = {Symposium (International) on Combustion},
  year    = {1953},
  volume  = {4},
  number  = {1},
  pages   = {498--501}
}

@incollection{powers1994,
  author    = {Powers, J. M.},
  title     = {Oblique detonations: theory and propulsion applications},
  booktitle = {Combustion in High-Speed Flows},
  editor    = {Buckmaster, J. and Jackson, T. L. and Kumar, A.},
  publisher = {Kluwer Academic},
  address   = {Dordrecht},
  year      = {1994},
  pages     = {345--371},
  doi       = {10.1007/978-94-011-1050-1_12}
}

@book{fickett2000,
  author    = {Fickett, W. and Davis, W. C.},
  title     = {Detonation: Theory and Experiment},
  publisher = {Dover},
  address   = {Mineola, NY},
  year      = {2000}
}

@article{emanuel2004,
  author  = {Emanuel, G. and Tuckness, D. G.},
  title   = {Steady, oblique, detonation waves},
  journal = {Shock Waves},
  year    = {2004},
  volume  = {13},
  number  = {6},
  pages   = {445--451},
  doi     = {10.1007/s00193-003-0222-1}
}

@article{li1994,
  author  = {Li, C. P. and Kailasanath, K. and Oran, E. S.},
  title   = {Detonation structures behind oblique shocks},
  journal = {Physics of Fluids},
  year    = {1994},
  volume  = {6},
  number  = {4},
  pages   = {1600--1611},
  doi     = {10.1063/1.868273}
}

@article{papalexandris2000,
  author  = {Papalexandris, M. V.},
  title   = {A numerical study of wedge-induced detonations},
  journal = {Combustion and Flame},
  year    = {2000},
  volume  = {120},
  number  = {4},
  pages   = {526--538},
  doi     = {10.1016/S0010-2180(99)00113-3}
}

@article{teng2014,
  author  = {Teng, H. H. and Jiang, Z. L. and Ng, H. D.},
  title   = {Numerical study on unstable surfaces of oblique detonations},
  journal = {Journal of Fluid Mechanics},
  year    = {2014},
  volume  = {744},
  pages   = {111--128},
  doi     = {10.1017/jfm.2014.78}
}

@article{nickalls1993,
  author  = {Nickalls, R. W. D.},
  title   = {A new approach to solving the cubic: {C}ardan's solution revealed},
  journal = {The Mathematical Gazette},
  year    = {1993},
  volume  = {77},
  number  = {480},
  pages   = {354--359},
  doi     = {10.2307/3619777}
}

@article{li2025criteria,
  author  = {Li, H. and Zhang, Z. and Wen, C.},
  title   = {Theoretical transition criteria in steady shock-incidence/detonation-reflection phenomenon},
  journal = {Combustion and Flame},
  year    = {2026},
  volume  = {284},
  pages   = {114627},
  doi     = {10.1016/j.combustflame.2025.114627}
}

@techreport{gordon1994cea,
  author      = {Gordon, S. and McBride, B. J.},
  title       = {Computer Program for Calculation of Complex Chemical Equilibrium Compositions and Applications. {P}art 1: Analysis},
  institution = {NASA},
  type        = {Reference Publication},
  number      = {1311},
  year        = {1994},
  url         = {https://ntrs.nasa.gov/citations/19950013764}
}

@article{zhang2022equilibrium,
  author  = {Zhang, Z. and Wen, C. and Zhang, W. and Liu, Y. and Jiang, Z.},
  title   = {A theoretical method for solving shock relations coupled with chemical equilibrium and its applications},
  journal = {Chinese Journal of Aeronautics},
  year    = {2022},
  volume  = {35},
  number  = {6},
  pages   = {47--62},
  doi     = {10.1016/j.cja.2021.08.021}
}

\end{document}